\documentclass[aps,prd,twocolumn,groupedaddress,nofootinbib,amssymb]{revtex4}

\def\theequation{\arabic{section}.\arabic{equation}} 
\usepackage{graphicx,bm,color}
\usepackage{amsmath} \usepackage{amssymb} \usepackage{amsfonts} 
\usepackage{wrapfig} \usepackage{cancel} \usepackage{url} 
\usepackage{amsmath} \usepackage{amssymb} \usepackage{amsfonts} 
\usepackage{graphicx,bm} \usepackage{dcolumn} \usepackage{color}
\usepackage{color,amsxtra} \usepackage{epsf}
\usepackage{amssymb} \usepackage{amsmath} \usepackage{enumerate} 
\usepackage{hhline} \usepackage{array} \usepackage{tabularx}

\newcommand{\be}{\begin{equation}} \newcommand{\ee}{\end{equation}} 
\newcommand{\bea}{\begin{eqnarray}} \newcommand{\eea}{\end{eqnarray}} 
\newcommand{\beaa}{\begin{eqnarray*}} \newcommand{\eeaa}{\end{eqnarray*}}

\newcommand{\nn}{\nonumber \\} 
\newcommand{\e}{\mathrm{e}}

\allowdisplaybreaks[4]

\begin{document} \tolerance=5000 
\def\theequation{\arabic{section}.\arabic{equation}}

\title{Searching for dynamical black holes in various theories of gravity}

\author{Shin'ichi~Nojiri$^{1,2}$, Sergei~D.~Odintsov$^{3,4,5,6}$ and 
Valerio~Faraoni$^{7}$, } 
\affiliation{ $^1$Department of Physics, Nagoya University, Nagoya 
464-8602, Japan\\ 
$^2$Kobayashi-Maskawa Institute for the Origin of Particles and the 
Universe, Nagoya University, Nagoya 464-8602, Japan\\ 
$^3$Instituci\'{o} Catalana de Recerca i Estudis 
Avan\c{c}ats (ICREA), Passeig Llu\'{i}s Companys, 23, 08010 Barcelona, 
Spain\\ 
$^4$ Institute of Space Sciences (ICE, CSIC) C. Can Magrans s/n, 
08193 Barcelona, Spain \\ 
$^5$ Institute of Space Sciences of Catalonia (IEEC), Barcelona, Spain\\ 
$^6$ Lab. Theor. Cosm., TUSUR, 634050 Tomsk and TSPU, 634061 Tomsk, 
Russia\\
$^7$Department of Physics and Astronomy, Bishop's University, 2600 College 
Street, Sherbrooke, Qu\'ebec, Canada J1M 1Z7 }

\begin{abstract}

We construct models of Einstein and $f(R)$ gravity with two scalar fields, 
which admit analytical solutions describing time-varying dynamical black 
holes. Their thermodynamics is investigated in the adiabatic approximation. 
In addition to the Misner-Sharp-Hernandez quasilocal mass, we provide 
time-dependent thermodynamical quantities, including the Hawking 
temperature, Helmholtz free energy, entropy, and thermodynamical energy.  
The latter does not always coincide with the Misner-Sharp-Hernandez mass at 
the horizon, although they coincide in the static limit. For 
Schwarzschild-type ({\em i.e.}, $g_{tt}g_{rr}=-1$) black holes in Einstein 
gravity, one of the two scalars is always a ghost with negative kinetic 
energy. We show that this ghost can be avoided in $f(R)$ gravity.

\end{abstract} \maketitle

\section{Introduction} \label{sec:1} \setcounter{equation}{0}

Black hole dynamics and thermodynamics, developed for stationary and 
asymptotically flat black holes ({\em e.g.}, 
\cite{Wald:1984rg,Poisson:2009pwt,Wald:1999vt}), are now mature fields of 
gravitational physics. However, no black hole is truly stationary or 
asymptotically flat. From the astrophysical point of view, realistic black 
holes interact with their environments in binary systems or in galaxies, 
through tidal forces, by accreting gas, and/or by emitting gravitational 
waves. From the purely theoretical point of view, black holes emit Hawking 
radiation losing energy, and are embedded in the expanding universe instead 
of being truly isolated, asymptotically flat, systems. Therefore, the 
ultimate theoretical description of black holes requires the consideration 
of {\em dynamical} solutions of the gravitational field equations. This is 
not a small step, both conceptually and computationally. Black holes are 
defined by their horizons: for stationary black holes, these are event 
horizons and null surfaces and their definition as a connected component of 
the boundary of the causal past of future null infinity \cite{Wald:1984rg} 
requires the knowledge of the entire causal structure (including the future 
history) of spacetime, which is summarized by saying that event horizons 
are teleological 
\cite{Nielsen:2008cr,Nielsen:2010gm,Booth:2005qc,Faraoni:2018xwo}. This 
traditional black hole concept becomes essentially useless for practical 
purposes in dynamical situations and is replaced by the more useful 
quasilocal definition of trapping surfaces and apparent or trapping 
horizons. Being defined quasilocally, apparent horizons require only the 
knowledge of a limited portion of spacetime 
\cite{Nielsen:2008cr,Nielsen:2010gm,Booth:2005qc,Faraoni:2018xwo,Vbook}. In 
spite of some disadvantages (most notably, a dependence on the spacetime 
foliation \cite{Wald:1991zz,Schnetter:2005ea}, and the fact that they are 
spacelike or timelike surfaces and that they can change their causal 
character during their dynamical evolution), apparent horizons are widely 
used in numerical simulations of the collisions of black holes with other 
compact objects that led to the interferometric discovery of gravitational 
waves \cite{Abbott:2016blz,Abbott:2016nmj,Abbott:2017vtc}. Given the need 
to predict in detail the waveforms of gravitational waves emitted by binary 
systems containing black holes, banks of templates for gravitational 
waveforms need to be built to separate signals from environmental and other 
noise in the laser interferometric detectors of gravitational waves. The 
notion of event horizon is of little use in the numerical study of fast 
astrophysical processes producing those gravitational waves. Instead, 
``black holes'' are routinely identified with outermost marginally trapped 
surfaces and apparent horizons in numerical research 
\cite{Nielsen:2008cr,Nielsen:2010gm,Booth:2005qc,Faraoni:2018xwo}. 
Numerical relativity uses apparent/trapping horizons, not event horizons.

Given this relatively recent paradigm shift, it would be useful to have a 
catalog of analytical solutions of general relativity (GR) and alternative 
theories of gravity describing dynamical black holes, and defined by their 
apparent horizons. Contrary to static and stationary black holes, very few 
dynamical black hole solutions are known in GR, and even fewer in 
alternative theories of gravity. Finding (and interpreting) new 
time-dependent exact solutions describing time-varying black holes is a 
rather difficult task already in GR. In principle, approaching this task in 
the context of alternative theories of gravity, which provide more 
gravitational degrees of freedom and, therefore, more flexibility should be 
easier, but this does not seem to be the case.

From the point of view of black hole thermodynamics, placing a black hole 
in a non-trivial environment changes its mass-energy, which plays the role 
of internal energy in the first law of black hole thermodynamics. This 
mass-energy is best defined quasi-locally. Here we set out to find 
analytical solutions of various theories of gravity and, for simplicity, we 
restrict ourselves to spherical symmetry, which has the added advantage 
that apparent horizons coincide in all spherical foliations 
\cite{Faraoni:2016xgy}. In this case the mass-energy of a spherical 
geometry, asymptotically flat or not, is the Misner-Sharp-Hernandez mass 
universally used in relativistic fluid dynamics and in studies of black 
hole collapse \cite{Misner:1964je,Hernandez:1966zia} and which coincides, 
in spherical symmetry, with the Hawking-Hayward quasilocal energy 
\cite{Hawking:1968qt,Hayward:1993ph}.

A few spherically symmetric solutions interpreted as genuine black holes 
embedded in Friedmann-Lema\^itre-Robertson-Walker (FLRW) universes are 
known in GR, beginning with the McVittie family\footnote{The well known 
Schwarzschild-deSitter/Kottler geometry, which is a special case of the 
McVittie, is instead locally static.} \cite{McVittie:1933zz}, which has 
been generalized \cite{Faraoni:2007es,Gao:2008jv} and has been the subject 
of much attention during the last decade 
\cite{Nolan:1993,Nolan:1998xs,Nolan:1999kk, 
Nolan:1999wf,Ferraris:1996ey,Kaloper:2010ec,Lake:2011ni,Landry:2012nv,Nandra:2011ug, 
Nandra:2011ui,Carrera:2009ve,daSilva:2015mja,daSilva:2012nh,Delliou:2013xra,
Faraoni:2014nba,Faraoni:2012gz,Gregoris:2019oxz}, also in the context of
Ho\v{r}ava-Lifschitz and Horndeski gravity \cite{Abdalla:2013ara, 
Afshordi:2014qaa}. The phenomenology of apparent horizons can be rather 
bizarre and provides several surprises, such as horizons appearing and 
disappearing in pairs, as in the Husain-Martinez-Nu\~nez solution (a black 
hole embedded in a FLRW universe sourced by a free scalar field) 
\cite{Husain:1994uj}. The non-rotating Thakurta solution \cite{Thakurta} is 
a late-time attractor of the generalized McVittie solutions 
\cite{Faraoni:2008tx}. Other solutions of Einstein's theory are less 
significant because they suffer from negative energy densities or 
instabilities in certain spacetime regions 
\cite{Sultana:2005tp,McClure:2006kg,McClure:2006aa,McClure:2007nk,McClure:2008zk}, 
while attempts to build new solutions have had mixed success 
\cite{Gao:2004cr,Nojiri:2017qvx,Finch:2011wz,Lindesay:2013iba,Lindesay:2010uv, 
Nozawa:2007vq,Pielahn:2011ra,Rodrigues:2015sja,Saida:2007ru,Sakai:2001gh, 
Sussman1985,Faraoni:2007gq,Faraoni:2009uy,Faraoni:2010rt,Nielsen:2011zy, 
C:2019zaw}.

Dynamical black hole solutions of alternative theories of gravity include 
the Clifton geometry in $f(R)=R^n$ gravity 
\cite{Clifton:2006ug,Faraoni:2009xb} and some members of the 
Clifton-Mota-Barrow family in Brans-Dicke gravity with a perfect fluid 
\cite{Clifton:2004st,Faraoni:2012sf}.

Since many alternative theories of gravity contain effective scalar degrees 
of freedom of gravitational origin, in spherical symmetry scalars either 
collapse to the Schwarzschild black hole \cite{Hawking:1972qk, 
Bekenstein:1996pn,Sotiriou:2011dz,Bhattacharya:2015iha} or tend to generate 
naked singularities or wormhole throats. It is not surprising that research 
in alternative theories produces these exotic objects more often than new 
black holes ({\em e.g.}, 
\cite{Fisher:1948yn,Bergmann:1957zza,Janis:1968zz,Buchdahl:1972sj, 
Wyman:1981bd,Bronnikov:1973fh,Campanelli:1993sm,Vanzo:2012zu,
Faraoni:2018mes,Fonarev:1994xq,Kastor:2016cqs, 
Faraoni:2017afs,Banijamali:2019gry,Faraoni:2017ecj}).
Here we dismiss these exotica and we look for dynamical black holes, 
characterized by time-varying apparent horizons, in various theories of 
gravity. We follow the notation and conventions of Ref.~\cite{Wald:1984rg}, 
using units in which the speed of light $c$ and Newton's constant $G$ are 
unity, while $\kappa^2 \equiv 8\pi G$.

\section{Spherically symmetric and time-dependent geometries} \label{sec:2} 
\setcounter{equation}{0}

The most general spherically symmetric and time-dependent line element in 
polar coordinates $\left( \tau, \rho, \vartheta, \varphi \right)$ is 
\begin{eqnarray} \label{G1} 
ds^2 &=& - \mathcal{A}(\tau,\rho) d\tau^2 + 2 
\mathcal{B} (\tau,\rho) d\tau d\rho
+ \mathcal{C}(\tau,\rho) d\rho^2 \nonumber\\ &\, & + \mathcal{D} 
  (\tau,\rho) \left( d\vartheta^2 + \sin^2\vartheta \, d\varphi^2 \right)  
  \, ,
\end{eqnarray} 
where the areal radius $r$ is defined by 
\begin{equation} 
\label{G2} r^2 \equiv \mathcal{D} (\tau,\rho) \, , 
\end{equation} 
and 
$\mathcal{D} (\tau,\rho)$ (as well as $\mathcal{A}$ and $ \mathcal{C}$) is 
necessarily positive to preserve the metric signature ${-}{+}{+}{+}$. In 
principle, Eq.~(\ref{G2}) can be solved for $\rho(\tau, r)$ (although, in 
practice, it may be difficult to invert explicitly the one-to-one relation 
$r=\sqrt{D\left( \tau, \rho \right)}$). In terms of the areal radius, the 
line element~(\ref{G1}) is rewritten as 
\begin{widetext} 
\begin{align} 
\label{G3} ds^2 =& \left[ - \mathcal{A}\left(\tau,\rho\left(\tau,r\right) 
\right) + 2 \mathcal{B} \left(\tau,\rho\left(\tau,r\right) \right) 
\frac{\partial\rho}{\partial\tau} \right] d\tau^2
+ 2 \mathcal{B} \left(\tau,\rho \left(\tau,r\right) \right) \frac{\partial 
  \rho}{\partial r} \, d\tau dr \nn
& + \mathcal{C}\left(\tau,\rho\left(\tau,r\right) \right) \left( 
   \frac{\partial \rho}{\partial r} \right)^2 dr^2
+ r^2 \left( d\vartheta^2 + \sin^2\vartheta \, d\varphi^2 \right) \, .  
\end{align} 
\end{widetext}  
In order to diagonalize this line element, we 
  introduce a new time coordinate $t$. Substituting $\tau=\tau(t,r)$ into 
  Eq.~(\ref{G3}), the line element becomes
\begin{widetext} 
\begin{align} 
  \label{G4} 
ds^2 =& \left[ - \mathcal{A}\left(\tau\left(t,r\right) , 
  \rho\left(\tau\left(t,r\right),r\right) \right)
+ 2 \mathcal{B} 
  \left(\tau\left(t,r\right),\rho\left(\tau\left(t,r\right),r\right)  
  \right)  
  \frac{\partial\rho\left(\tau\left(t,r\right),r\right)}{\partial\tau} 
  \right] \left( \frac{\partial \tau\left(t,r\right)}{\partial t} \right)^2 
  dt^2 \nn 
& + 2 \left[ \mathcal{B} 
  \left(\tau\left(t,r\right),\rho\left(\tau\left(t,r\right),r\right)  
  \right) \frac{\partial \rho\left(\tau\left(t,r\right),r\right)}{\partial 
  r} \, \frac{\partial \tau\left(t,r\right)}{\partial t} \right. \nn
& \left. + \left\{ - \mathcal{A}\left(\tau\left(t,r\right), 
   \rho\left(\tau\left(t,r\right),r\right) \right)
+ 2 \mathcal{B} 
  \left(\tau\left(t,r\right),\rho\left(\tau\left(t,r\right),r\right)  
  \right)  
  \frac{\partial\rho\left(\tau\left(t,r\right),r\right)}{\partial\tau} 
  \right\} \right. \nn
& \left. \qquad \times \, \frac{\partial \tau\left(t,r\right)}{\partial t}
\, \frac{\partial \tau\left(t,r\right)}{\partial r} \right] dt dr \nn
& + \left[ \mathcal{C}\left(\tau,\rho\left(\tau,r\right) \right) \left( 
   \frac{\partial \rho}{\partial r} \right)^2
+ \mathcal{B} \left( \tau\left(t,r\right),\rho\left( t,r\right),r\right)
   \frac{\partial \rho\left(\tau\left(t,r\right),r\right)}{\partial r} \, 
  \frac{\partial \tau\left(t,r\right)}{\partial r} \right. \nn
& + \left\{ - \mathcal{A}\left(\tau\left(t,r\right), 
   \rho\left(\tau\left(t,r\right),r\right) \right)
+ 2 \mathcal{B} 
  \left(\tau\left(t,r\right),\rho\left(\tau\left(t,r\right),r\right)  
  \right)  
  \frac{\partial\rho\left(\tau\left(t,r\right),r\right)}{\partial\tau} 
  \right\} \nn
& \left. \qquad \times \left( \frac{\partial \tau\left(t,r\right)}{\partial
r} \right)^2 \right] dr^2 + r^2 \left( d\vartheta^2 + \sin^2\vartheta \, 
d\varphi^2 \right) \, . 
\end{align}  
By choosing the new time coordinate $t$ 
so that 
\begin{align} 
\label{G5} 
0=& \, \mathcal{B} \left( 
\tau\left(t,r\right),\rho\left( \tau\left(t,r\right),r\right) \right) 
\frac{\partial \rho \left( \tau\left(t,r\right),r\right)}{\partial r} \, 
\frac{\partial \tau\left(t,r\right)}{\partial t} \nn
& + \left[ - \mathcal{A}\left(\tau\left(t,r\right),
\rho\left( \tau\left(t,r\right),r \right) \right)
+ 2 \mathcal{B} 
  \left(\tau\left(t,r\right),\rho\left(\tau\left(t,r\right),r\right)  
  \right)  
  \frac{\partial\rho\left(\tau\left(t,r\right),r\right)}{\partial\tau} 
  \right]
\frac{\partial \tau\left(t,r\right)}{\partial t} \, \frac{\partial 
\tau\left(t,r\right)}{\partial r} \, , 
\end{align} 
\end{widetext}  
the line 
element assumes the diagonal form 
\begin{equation} 
\label{GBiv} 
ds^2 = - 
\e^{2\nu (t,r)} dt^2 + \e^{2\lambda (t,r)} dr^2
+ r^2 \left( d\vartheta^2 + \sin^2\vartheta \, d\varphi^2 \right)\, , 
  \end{equation} where \begin{widetext} \begin{align}
- \e^{2\nu (t,r)} \equiv & \left\{
- \mathcal{A}\left( \tau\left(t,r\right), 
  \rho\left(\tau\left(t,r\right),r\right) \right)
+ 2 \mathcal{B} \left(\tau\left(t,r\right), 
  \rho\left(\tau\left(t,r\right),r\right) \right)  
  \frac{\partial\rho\left(\tau\left(t,r\right),r\right)}{\partial\tau} 
  \right\} 
\left( \frac{\partial \tau\left(t,r\right)}{\partial t} \right)^2 \, , \\
\e^{2\lambda (t,r)} \equiv & \, 
\mathcal{C}\left(\tau,\rho\left(\tau,r\right) \right) \left( \frac{\partial 
\rho}{\partial r} \right)^2
+ \mathcal{B} 
  \left(\tau\left(t,r\right),\rho\left(\tau\left(t,r\right),r\right)  
  \right) \frac{\partial \rho\left(\tau\left(t,r\right),r\right)}{\partial 
  r} \frac{\partial \tau\left(t,r\right)}{\partial r} \nn
& + \left[ - \mathcal{A}\left(\tau\left(t,r\right),
\rho\left(\tau\left(t,r\right),r\right) \right)
+ 2 \mathcal{B} 
  \left(\tau\left(t,r\right),\rho\left(\tau\left(t,r\right),r\right)  
  \right)  
  \frac{\partial\rho\left(\tau\left(t,r\right),r\right)}{\partial\tau} 
  \right] \left( \frac{\partial \tau\left(t,r\right)}{\partial r} \right)^2 
  \, . \end{align} \end{widetext} We define the metric $\bar{g}_{ij}$ of 
  the unit 2-sphere by $\sum_{i,j=1,2} \bar{g}_{ij} dx^i dx^j = 
  d\vartheta^2 + \sin^2\vartheta \, d\varphi^2$. For the 
  metric~(\ref{GBiv}), the only non-vanishing connection coefficients are
\begin{align} \label{GBv0}
&\Gamma^t_{tt}=\dot\nu \, , \quad \Gamma^r_{tt}
= \e^{-2(\lambda - \nu)}\nu' \, , \quad \Gamma^t_{tr}=\Gamma^t_{rt}=\nu'\, 
, \nn
& \Gamma^t_{rr} = \e^{2\lambda - 2\nu}\dot\lambda \, , \quad
\Gamma^r_{tr} = \Gamma^r_{rt} = \dot\lambda \, , \nn
& \Gamma^r_{rr}=\lambda'\, , \Gamma^i_{jk} = \bar{\Gamma} ^i_{jk}\, ,\quad
\Gamma^r_{ij}=-\e^{-2\lambda}r \bar{g}_{ij} \, , \nn
& \Gamma^i_{rj}=\Gamma^i_{jr}=\frac{1}{r} \, \delta^i_{\ j}\,,
\end{align} where $\bar{ \Gamma}^i_{jk}$ is the metric connection of 
$\bar{g}_{ij}$, while an overdot and a prime denote differentiation with 
respect to $t$ and $r$, respectively. Using the expression of the Riemann 
tensor \cite{Wald:1984rg} \begin{equation} \label{Riemann} {R^\lambda}_{\ 
\mu\rho\nu} =\Gamma^\lambda_{\mu\nu,\rho} -\Gamma^\lambda_{\mu\rho,\nu}
+ \Gamma^\eta_{\mu\nu}\Gamma^\lambda_{\rho\eta}
- \Gamma^\eta_{\mu\rho}\Gamma^\lambda_{\nu\eta} \,, \end{equation} one 
  finds \begin{widetext} \begin{align} \label{curvatures} R_{rtrt} = & - 
  \e^{2\lambda} \left[ \ddot\lambda
+ \left( \dot\lambda - \dot\nu \right) \dot\lambda \right]
+ \e^{2\nu}\left[ \nu'' + \left(\nu' - \lambda'\right)\nu' \right] \, ,\nn 
  R_{titj} =& \, r\nu' \, \e^{2(\nu - \lambda)} \bar{g}_{ij} \, ,\nn
R_{rirj} =& \, \lambda' r \bar{ g}_{ij} \, ,\quad {R_{tirj}= \dot\lambda r 
\bar{ g}_{ij} } \, , \quad R_{ijkl} = \left( 1 - \e^{-2\lambda}\right) r^2 
\left(\bar{g}_{ik} \bar{g}_{jl} - \bar{g}_{il} \bar{g}_{jk} \right)\, ,\nn
R_{tt} =& - \left[ \ddot\lambda
+ \left( \dot\lambda - \dot\nu \right) \dot\lambda \right]
+ \e^{2\left(\nu - \lambda\right)} \left[ \nu'' + \left(\nu' - 
  \lambda'\right)\nu' + \frac{2\nu'}{r}\right] \, ,\nn
R_{rr} =& \, \e^{-2\left( \nu - \lambda \right)} \left[ \ddot\lambda
+ \left( \dot\lambda - \dot\nu \right) \dot\lambda \right]
- \left[ \nu'' + \left(\nu' - \lambda'\right)\nu' \right]
+ \frac{2 \lambda'}{r} \, ,\nn 
  , \quad R_{ij} = \left[ 1 + \left\{ - 1 - r \left(\nu' - \lambda' 
  \right)\right\} \e^{-2\lambda}\right] \bar{g}_{ij}\ , \nn
R=& \, 2 \, \e^{-2 \nu} \left[ \ddot\lambda
+ \left( \dot\lambda - \dot\nu \right) \dot\lambda \right] \nn 
& + \e^{-2\lambda}\left[ - 2\nu'' - 2\left(\nu'
 - \lambda'\right)\nu' - \frac{4\left(\nu'
 - \lambda'\right)}{r} + \frac{2\e^{2\lambda} - 2}{r^2} \right] \, .  
   \end{align}\end{widetext}

Any spherically symmetric metric can be recast in the Abreu-Nielsen-Visser 
gauge \cite{Nielsen:2005af,Abreu:2010ru} 
\begin{eqnarray} 
ds^2 &=& -  \mathrm{e}^{-2\Phi(t,r)} 
\left(1-\frac{2M_{\mathrm{MSH}}(t,r)}{r} \right)  dt^2 \nn
&\, & +\frac{dr^2}{1-\frac{2M_\mathrm{MSH}(t,r)}{r}} \nn 
&\, & + r^2
\left(d \vartheta^2 +\sin^2 \vartheta \, d\varphi^2 \right) 
\,,\label{AbreuVissergauge} 
\end{eqnarray} 
where $r$ is the areal radius 
and $M_{\mathrm{MSH}}$ is the Misner-Sharp-Hernandez mass defined in 
spherical symmetry by \cite{Misner:1964je,Hernandez:1966zia} 
\be 
1-\frac{2M_{\mathrm{MSH}}}{r} = \nabla^cr \nabla_c r 
\ee 
which, in the 
gauge~(\ref{AbreuVissergauge}), reduces to $g^{rr}=1-2M_{\mathrm{MSH}}/r$. 
(Although this definition was originally given in GR 
\cite{Misner:1964je,Hernandez:1966zia}, it applies to other theories of 
gravity as well, but its role in the relevant equations may change 
\cite{Faraoni:2015sja,Giusti:2020uuy}.) The more general Hawking-Hayward 
quasilocal mass \cite{Hawking:1968qt,Hayward:1993ph} reduces to the 
Misner-Sharp-Hernandez mass in spherical symmetry \cite{Hayward:1994bu} and 
is not restricted to asymptotically flat spacetimes (however, in the 
asymptotically flat case, the Hawking-Hayward/Misner-Sharp-Hernandez mass 
computed at spatial infinity reduces to the Arnowitt-Deser-Misner (ADM) 
mass \cite{Arnowitt:1962hi}).

By comparing the Abreu-Nielsen-Visser gauge~(\ref{AbreuVissergauge}) with 
Eq.~(\ref{GBiv}), one has the correspondence \begin{align} 
\mathrm{e}^{2\nu} =& \, \mathrm{e}^{-2\Phi} \left( 1
 - \frac{2M_{\mathrm{MSH}} }{r}\right) = \mathrm{e}^{- 2 \left(\lambda + 
   \Phi \right)} \,,\\
\mathrm{e}^{2\lambda} =& \left( 1-\frac{2M_{\mathrm{MSH}} }{r} \right)^{-1} 
\,. \end{align} We are interested in apparent horizons, which can be 
dynamical, and are located by the roots of $\nabla^c r\nabla_c r=0$, or 
$r_{\mathrm{AH}}=2M_{\mathrm{MSH}}( r_{\mathrm{AH}})$ 
\cite{Misner:1964je,Hernandez:1966zia,Abreu:2010ru,Vbook}.

\section{Spherically symmetric and time-dependent solutions of 
Einstein-two-scalar models} \label{sec:3} \setcounter{equation}{0}

Let us first consider GR with two scalar fields $\phi$ and $\chi$ as the 
matter source, as described by the action \begin{widetext} \begin{align} 
\label{I8} S_{( \mathrm{GR} \phi\chi)} = \int d^4 x \sqrt{-g} & \left[ 
\frac{R}{2\kappa^2}
 - \frac{1}{2} \, A (\phi,\chi) \partial_\mu \phi \partial^\mu \phi
 - B (\phi,\chi) \, \partial_\mu \phi \partial^\mu \chi \right. \nn 
  & \left. \quad - \frac{1}{2} \, C (\phi,\chi) \partial_\mu \chi
   \partial^\mu \chi - V (\phi,\chi)\right] \, , \end{align} \end{widetext} 
where $g$ is the determinant of the metric tensor $g_{\mu\nu}$, $R$ is the 
Ricci scalar, $V(\phi, \chi)$ is the potential of the scalar doublet, and 
the coefficients $A$, $B$, and $C$ depend on the scalars. Scenarios 
with two or more scalar fields have been studied in the literature many 
times, especially in relation with multiple scalar field inflation ({\em 
e.g.}, \cite{GarciaBellido:1995qq, Coley:1999mj, Wands:2002bn, Do:2011zza, 
Bamba:2015uxa, Elizalde:2008yf}), but there is no specific relation 
between those models 
and the theories that we use in the present manuscript. The matter 
energy-momentum tensor is 
\begin{widetext} 
\begin{align} \label{I9} 
T^{(\phi\chi)}_{\mu\nu} =& g_{\mu\nu} \left[
 - \frac{1}{2}\, A (\phi,\chi) \partial_\rho \phi \partial^\rho \phi
 - B (\phi,\chi) \partial_\rho \phi \partial^\rho \chi
 - \frac{1}{2} \, C (\phi,\chi) \partial_\rho \chi \partial^\rho \chi - V 
   (\phi,\chi)\right] \nn
& + A (\phi,\chi) \partial_\mu \phi \partial_\nu \phi
+ B (\phi,\chi) \left( \partial_\mu \phi \partial_\nu \chi
+ \partial_\nu \phi \partial_\mu \chi \right)
+ C (\phi,\chi) \partial_\mu \chi \partial_\nu \chi \, , \end{align} and 
  the contracted Bianchi identities read \begin{align} \label{I10} 0 =& \, 
  \frac{A_\phi}{2}\, \partial_\mu \phi \partial^\mu \phi
+ A \nabla^\mu \partial_\mu \phi + A_\chi \partial_\mu \phi \partial^\mu 
  \chi + \left( B_\chi - \frac{1}{2} \, C_\phi \right)\partial_\mu \chi 
  \partial^\mu \chi + B \nabla^\mu \partial_\mu \chi - V_\phi \, ,\\
\label{I10b} 0 =& \left( - \frac{1}{2} \, A_\chi + B_\phi \right) 
\partial_\mu \phi \partial^\mu \phi + B \nabla^\mu \partial_\mu \phi
+ \frac{1}{2} \, C_\chi \partial_\mu \chi \partial^\mu \chi + C \nabla^\mu 
  \partial_\mu \chi + C_\phi \partial_\mu \phi \partial^\mu \chi -
V_\chi\, , \end{align} \end{widetext} where $A_\phi \equiv \partial 
A(\phi,\chi)/\partial \phi$, {\em etc}. We now identify \begin{equation} 
\label{TSBH1} \phi=t\, , \quad \chi=r\, . \end{equation} This assumption 
does not lead to a loss of generality for the following reason: for the 
spherically symmetric solutions~(\ref{GBiv}) of the theory~(\ref{I8}), in 
general $\phi$ and $\chi$ depend on both coordinates $t$ and $r$. Given a 
solution, the $t$- and $r$-dependence of $\phi$ and $\chi$ can be 
determined and $\phi$ and $\chi$ are then given by specific functions 
$\phi(t,r)$, $\chi(t,r)$. On spacetime regions where these relations are 
one-to-one, and provided that $\partial_{\mu} \phi $ is timelike and 
$\partial_{\mu} \chi$ is spacelike, one can invert these specific 
functional forms and redefine the scalar fields to replace $t$ and $r$ with 
new scalar fields, say, $\bar{\phi}$ and $\bar{\chi}$ with $\phi(t,r)\to 
\phi(\bar{\phi}, \bar{\chi})$ and $\chi(t,r) \to \chi(\bar{\phi}, 
\bar{\chi})$. We can then identify the new fields with $t$ and $r$ as 
in~(\ref{TSBH1}), $\bar{\phi}\to \phi=t$ and $\bar{\chi} \to \chi= r$. The 
change of variables $\left( \phi , \chi\right) \rightarrow \left( 
\bar{\phi} , \bar{\chi} \right)$ can then be absorbed into redefinitions of 
$A$, $B$, $C$, and $V$ in the action~(\ref{I8}). Therefore, the 
assumption~(\ref{TSBH1}) does not lead to loss of generality.

Proceeding, the $\left( t,t \right)$, $\left(r,r\right)$, 
$\left(i,j\right)$, and $\left(t,r \right)$ components of the Einstein 
equation assume, respectively, the form \begin{widetext} \begin{align} 
\label{TSBH2}
& \frac{\e^{2\left(\nu - \lambda\right)}}{\kappa^2}
\left( \frac{2\lambda'}{r} + \frac{\e^{2\lambda} - 1}{r^2} \right) = - 
\e^{2\nu} \left( - \frac{A}{2} \, \e^{-2\nu} - \frac{C}{2} \, 
\e^{-2\lambda} - V \right) \, ,\\
& \frac{1}{\kappa^2} \left( \frac{2\nu'}{r} - \frac{\e^{2\lambda}
 - 1}{r^2} \right) = \e^{2\lambda} \left( \frac{A}{2} \, \e^{-2\nu}
+ \frac{C}{2} \, \e^{-2\lambda} - V \right) \, ,\\ 
&  \frac{1}{\kappa^2} \left[ - \e^{-2 \nu} \left\{ \ddot\lambda
+ \left( \dot\lambda - \dot\nu \right) \dot\lambda \right\}
+ \e^{-2\lambda}\left( r \left(\nu' - \lambda' \right)
+ r^2 \nu'' + r^2 \left( \nu' - \lambda' \right) \nu' \right) \right] \nn
& \qquad = r^2 \left( \frac{A}{2} \, \e^{-2\nu} - \frac{C}{2} \,
\e^{-2\lambda} - V \right) \, ,\\
\label{TSBH5}
& \frac{2\dot\lambda}{\kappa^2 r} = B \,.
\end{align} \end{widetext} Due to spherical symmetry, 
the other components of the Einstein equation are trivially satisfied. 
Equations~(\ref{TSBH2})-(\ref{TSBH5}) can be solved with respect to $A$, 
$B$, $C$, and $V$, obtaining the inverse relations \begin{widetext} 
\begin{align} \label{TSBH6} A=& \,\frac{1}{\kappa^2} \left[ - \left\{ 
\ddot\lambda
+ \left( \dot\lambda - \dot\nu \right) \dot\lambda \right\}
+ \e^{2\left(\nu - \lambda\right)} \left( \frac{\e^{2\lambda} - 1}{r^2}
+ \frac{\nu' + \lambda'}{r} + \nu'' + \left( \nu' - \lambda' \right) \nu' 
  \right) \right] \, , \\
\label{TSBH7} B=& \, \frac{2\dot\lambda}{\kappa^2 r} \, , \\
\label{TSBH8} C=& \, \frac{1}{\kappa^2} \left[ \frac{\e^{-2\left(\nu - 
\lambda \right)}}{r^2}\left\{ \ddot\lambda
+ \left( \dot\lambda - \dot\nu \right) \dot\lambda \right\}
 - \frac{\e^{2\lambda} - 1}{r^2}
+ \frac{\nu' + \lambda'}{r} - \nu'' - \left( \nu' - \lambda' \right) \nu' 
  \right] \, , \\
\label{TSBH9} V=& \, \frac{\e^{-2\lambda}}{2\kappa^2} \left[ \frac{2\left( 
\lambda' - \nu'\right) }{r}
+ \frac{2 \left(\e^{2\lambda} - 1\right) }{r^2} \right] \, . \end{align} 
  \end{widetext} The right hand sides of Eqs.~(\ref{TSBH6})-(\ref{TSBH9})  
  are functions of $t$ and $r$. Replacing $\left( t, r\right) $ with 
  $\left( \phi , \chi \right) $ in these right hand sides, we obtain $A$, 
  $B$, $C$, and $V$ as functions of $\left( \phi , \chi \right)$.  
  Conversely, if we prescribe $A$, $B$, $C$, and $V$, the model has as 
  solutions spherically symmetric configurations~(\ref{GBiv}) corresponding 
  to arbitrary functions $\nu$ and $\lambda$.

We may also consider solutions for which $\lambda=-\nu$, as in the 
Schwarzschild geometry. In a spherical spacetime in which the line element 
is written using the areal radius $r$ and $g_{tt}g_{rr}=-1$, $r$ is an 
affine parameter along radial null geodesics \cite{Jacobson:2007tj}. 
Furthermore, such a spacetime enjoys special algebraic properties 
(\cite{Jacobson:2007tj}, see also \cite{BondiKilmister,Dadhich:2015wgx}): 
the double projection $R_{\mu\nu} l^{\mu} l^{\nu}$ of the Ricci tensor onto 
radial null vectors $l^\alpha$ vanishes identically \cite{Jacobson:2007tj}. 
Alternatively, the restriction of the Ricci tensor to the $\left(t, r 
\right)$ submanifold is proportional to the restriction of the metric 
$g_{\mu\nu}$ to this subspace \cite{Jacobson:2007tj}. Many analytical 
solutions of interest in GR or in alternative gravities satisfy the 
condition $g_{tt}g_{rr}=-1$, including vacuum solutions, electrovacuum 
solutions with the Maxwell or with nonlinear Born-Infeld electrodynamics 
\cite{Jacobson:2007tj}, and the string hedgehog global monopole 
\cite{Barriola:1989hx,Guendelman:1991qb}, also in higher dimension.

Under the condition $\lambda=-\nu$, Eqs.~(\ref{TSBH6})-(\ref{TSBH9}) 
simplify to \begin{widetext} \begin{align} \label{TSBH6BB} A=& - r^2 
\e^{4\nu} C = \frac{1}{\kappa^2} \left[ - \frac{ \e^{2\nu} }{2}\, \frac{d^2 
\left( \e^{-2\nu} \right)}{dt^2} + \e^{4\nu} \left( \frac{\e^{- 2\nu}
 - 1}{r^2}
+ \frac{ \e^{-2\nu}}{2} \, \frac{d^2 \left( \e^{2\nu} \right)}{dr^2} 
  \right) \right] \, ,\\
\label{TSBH7BB} B=& \, \frac{\e^{2\nu}}{\kappa^2 r} \, \frac{d\left( 
\e^{-2\nu} \right)}{dt} \, ,\\
\label{TSBH9BB} V= & \, \frac{\e^{2\nu}}{2\kappa^2} \left( - \frac{2 
\e^{-2\nu}}{r} \frac{d \left( \e^{2\nu} \right) }{dr} + \frac{2 \left(\e^{- 
2\nu}
 - 1\right) }{r^2} \right) \, . \end{align} \end{widetext} Since $AC = - 
   r^2 \e^{4\nu} C^2 <0$, either $\phi$ or $\chi$ has negative kinetic 
   energy in the action~(\ref{I8}) and a ghost is always present in this 
   case.

We may further restrict ourselves the static case, $\nu=\nu(r)$ and 
$\lambda=\lambda(r)$; then, Eqs.~(\ref{TSBH6BB}) and (\ref{TSBH7BB}) tell 
us that $A$, $B$, and $C$ depend only on $\chi$ and not on $\phi$. However, 
even in the static case, we still need $\phi$ to obtain a model 
corresponding to arbitrary $\nu$ and $\lambda$.

\subsection{Ansatz~1}

As an example, consider the ansatz \begin{equation} \label{E1} 
\e^{2\nu}=\e^{-2\lambda}= 1 - \frac{2M(t)}{r} \, , \end{equation} where the 
Misner-Sharp-Hernandez mass $M(t)$ is positive and depends only on time 
(clearly, a negative $M_{\mathrm{MSH}}$ signals a violation of the energy 
conditions and makes apparent horizons impossible). The Ricci curvature is 
\begin{equation} \label{TSBHFR9} R= \left( 1 - \frac{2M}{r} \right)^{-1} 
\left[ \frac{2 \ddot M}{r}
+ \frac{\left( \frac{2 \dot M}{r} \right)^2}{1 - \frac{2 M}{r}} \right] 
  \end{equation} and it vanishes if $M$ is constant, in which case the 
  geometry degenerates into the Schwarzschild one. There is only one 
  apparent horizon, with areal radius $r_{\mathrm{AH}}(t)=2M(t)$. This 
  horizon is dynamical and, since it is a single root of the equation 
  $\nabla^cr \nabla_c r=0$, it is a black hole horizon. (A curvature 
  singularity corresponds to $ 2M/r =1$ if $\frac{2 \ddot M}{r} \left( 1 - 
  \frac{2 M}{r} \right)
+ \left( \frac{2 \dot M}{r} \right)^2$ vanishes as $\mathcal{O} \left( 
  \left( 1 - \frac{2 M}{r} \right)^2 \right)$ when $2 M/r =1$, which is 
  impossible.)

To discuss the thermodynamics of the spacetime~(\ref{E1}), we assume that 
the time evolution is sufficiently slow ($\dot{M}\ll 1$) and we 
regard the time $t$ as a constant for the purposes of thermodynamics. This 
adiabatic approximation is necessary in order to avoid dealing with full 
non-equilibrium thermodynamics, which is essentially unknown for dynamical 
black holes \cite{Vbook}. It is present (although often not made 
explicit) in the tunneling formalism for dynamical black holes (see 
\cite{Vanzo:2011wq} for a review). In practice, it amounts to require that 
the horizon moves ``slowly'', or that the time derivatives of the functions 
used in the thermodynamical calculations are much smaller than the 
corresponding spatial derivatives (the Kodama time is used in the tunneling 
method, but other gauges are in principle possible). In the adiabatic 
regime, the temperature is 
\begin{equation} 
\label{T1} 
T=\frac{1}{8 \pi 
M(t)}\, , 
\end{equation} 
and the Bekenstein-Hawking entropy reads 
\begin{equation} 
\label{T2} 
\mathcal{S} = \frac{4\pi \left[ 
r_{\mathrm{AH}}(t) \right]^2}{4} = \frac{1}{16\pi T^2} \,. 
\end{equation} 
Then, since $dF/dT=- \mathcal{S}$, where $F$ is the free energy, we have 
\begin{equation} \label{T3} F= \frac{1}{16\pi T} \end{equation} by choosing 
the integration constant so that $F\rightarrow 0 $ in the formal limit 
$T\to \infty$ (or $M(t)\to 0$). The thermal energy is \begin{equation} 
\label{T4} E=F+T\mathcal{S} = \frac{1}{8\pi T} =M(t) \, . \end{equation} If 
$M$ is constant, the expressions of $T$, $\mathcal{S}$, and $E$ reduce to 
those of the Schwarzschild black hole. We also find 
\begin{align} 
\label{E3} 
A=& - \frac{1}{2\kappa^2} \left[ \frac{2 M''(\phi)}{\chi} + 
\frac{\left( \frac{2 M'(\phi)}{\chi} \right)^2} {1 - \frac{2 
M(\phi)}{\chi}}\right] \,\nn B=& {\, \frac{\frac{2 M'(\phi)}{\chi}} 
{\kappa^2 r\left( 1 - \frac{2 M(\phi)}{\chi} \right)}
} \, , 
\nn 
C=& \, \frac{1}{2\kappa^2\chi^2 \left( 1 - \frac{2 
M(\phi)}{\chi} \right)^2} \left[ \frac{2 M''(\phi)}{\chi} + \frac{\left( 
 \frac{2 M'(\phi)}{\chi} \right)^2} {1 - \frac{2 M(\phi)}{\chi}}\right]  
\,,\nn 
V=& \, 0 \, . 
\end{align} 
Since $AC<0$, one of the two free scalars $\phi$ or $\chi$ is always a 
ghost, which makes the theory physically inconsistent.

\subsection{Ansatz~2}

It is sometimes possible to find exact solutions by separating the time and 
space dependence in the metric coefficients (as done, for example, in 
\cite{Clifton:2004st,Clifton:2006ug}). As another example, consider the 
ansatz \begin{equation} \label{Ex2a} \e^{2\nu} \, = \, \e^{-2\lambda}= 
\left( 1 - \frac{r_0}{r} \right) \frac{t_0}{t} \,, \end{equation} where 
$r_0$ and $t_0$ are positive constants. The Misner-Sharp-Hernandez 
quasilocal mass is \be M_{\mathrm{MSH}}(t,r)= \frac{r}{2}\left[ 
1-\left(1-\frac{r_0}{r} \right) \frac{t_0}{t} \right] \ee and there is only 
one apparent horizon located by 
$r_{\mathrm{AH}}=2M_{\mathrm{MSH}}\left(t,r_{\mathrm{AH}}\right)$, which 
gives $r=r_0$.  Since this is a single root, we have a black hole apparent 
horizon. Remarkably, although the metric and the Misner-Sharp-Hernandez 
mass\footnote{The Misner-Sharp-Hernandez mass at the horizon 
$M_{\mathrm{MSH}} (r_0)= r_0/2$ is time-independent.} (of a generic sphere 
of radius $r$) are time-dependent, this apparent horizon has constant 
(areal) radius. This horizon is a null surface and an event horizon. In 
fact, if $ h(r) \equiv r-r_0$, the horizon is the surface $h(r)=0$ with 
normal \be N_{\mu}= \left. \nabla_{\mu} h \right|_{r=r_0}= \delta_{\mu 1} 
\,, \ee which is a null vector: \be g_{\mu\nu} N^{\mu} N^{\nu} = \left. 
g^{11}\right|_{r=r_0}= \left. \left( 1-\frac{r_0}{r} \right) \frac{t_0}{t} 
\right|_{r=r_0}=0 \,. \ee We have, therefore, a static event horizon in a 
non-stationary spacetime: the question of whether such a horizon can exist 
was posed, and approached perturbatively, in Ref.~\cite{Davidson:2012si}. 
Furthermore, since there is no timelike Killing vector, this event horizon 
is not a Killing horizon, which is of interest in relation with the strong 
rigidity theorem \cite{Hawking:1971vc,Hawking:1973uf} stating that the 
event horizon of a stationary black hole spacetime is a Killing horizon. 
This theorem is not violated here because it requires the matter 
stress-energy tensor to satisfy the weak energy condition while, as we 
shall see shortly, one of the two scalar fields is a phantom and violates 
it.

The Ricci curvature is \begin{equation} \label{Ex2c} R= \frac{2}{r^2} 
\left( 1 - \frac{t_0}{t} \right) \,. \end{equation} This geometry has no 
spacetime singularities except for the usual one at $r=0$ and the Big Bang 
at $t=0$.

With regard to the thermodynamics of the spacetime~(\ref{Ex2a}), we assume 
again an adiabatic regime (which is possibile when $t$ is not close 
to zero and the metric changes relatively slowly); then the temperature 
and the Bekenstein-Hawking entropy are (see Appendix~\ref{AppendixA} for 
details) 
\begin{equation} 
\label{T1e2} 
T=\frac{t_0}{4 \pi t r_0}\, , 
\end{equation} 
and 
\begin{equation} 
\label{T2e2} 
\mathcal{S} = \frac{4\pi 
r_0^2}{4} \, . 
\end{equation}

Even though the apparent horizon is static, together with its area 
$\mathcal{A}_{\mathrm{AH}}$ and the entropy 
$\mathcal{S}=\mathcal{A}_{\mathrm{AH}}/4$, the Hawking temperature 
decreases monotonically with time, which is interpreted as an effect of the 
accretion of the scalar doublet onto the black hole. Indeed, 
$\dot{\lambda}\neq 0$ and, therefore, $B\neq 0$ (as follows from 
Eq.~(\ref{I9})) and the component $T_{tr}^{(\phi\chi)} $ of the 
stress-energy tensor of the scalar doublet is non-vanishing, signalling a 
non-zero radial energy flux onto the black hole. It is noteworthy that, 
although the metric varies in time (which causes the Hawking temperature to 
vary according to Eq.~(\ref{T1e2})), the apparent horizon does not change 
its location. We are not aware of a similar occurrence in previous 
literature.

Eliminating $r_0$ with Eq.~(\ref{T1e2}) gives \begin{equation} 
\label{T2Be2} \mathcal{S} = \frac{t_0^2}{4 t^2 T^2} \,, \end{equation} and 
integrating $dF/dT=- \mathcal{S}$ yields the Helmholtz free energy 
\begin{equation} \label{T3e2} F= \frac{t_0^2}{4 t^2 T} \end{equation} and 
the thermal energy \begin{equation} \label{T4ex2} E=F+T\mathcal{S}= 
\frac{t_0^2}{2 t^2 T} \,, \end{equation} and we obtain thermodynamical 
quantities with non-trivial time dependence that obey the first law 
\begin{equation} \label{T5ex2} dE= - \frac{t_0^2}{2 t^2 T^2} \, dT= T dS = 
dQ \, , \end{equation} where $dQ$ is the heat transferred across the 
horizon.

We also find \begin{align} \label{TSBH6Ex2} A =& - \left( 1 - 
\frac{r_0}{\chi} \right)^2 \frac{t_0^2 \,\chi^2}{\phi^2} \, , \quad \\
\label{TSBH7Ex2} B=& \, \frac{1}{\kappa^2 rt} \, ,\\
C = & \, \frac{1}{\kappa^2} \left( 1 - \frac{r_0}{\chi} \right) 
\frac{t_0^2}{\phi^2} \left( \frac{\phi}{t_0 \chi^2} - \frac{1}{\chi^2} 
\right) \, ,\\
\label{TSBH9V} V=& \, \frac{\e^{-2\lambda}}{2\kappa^2} \left[ \frac{2\left( 
\lambda'
 - \nu'\right) }{r} + \frac{2 \left(\e^{2\lambda} - 1\right) }{r^2} \right] 
   \, . \end{align} Again, it is $AC<0$ and $\phi$ or $\chi$ is always a 
   ghost.

\subsection{Ansatz~3}

Our last example choice is \begin{equation} \label{Ex3A1} \e^{2\nu} \, = \, 
\e^{-2\lambda} \, = \frac{1 - \frac{r_0}{r} }{ 1
+ \frac{t r_0}{t_0 \, r} } \, . \end{equation} The Misner-Sharp-Hernandez 
  quasilocal mass is \be M_{\mathrm{MSH}}(t,r) = \frac{r_0}{2} \left( 
  \frac{1+\frac{t}{t_0}}{1+\frac{tr_0}{t_0r }}\right)
\ee and the apparent horizons are located by $r=2M_{\mathrm{MSH}}$, which 
yields the only root $r=r_0$. This is a single root and the radius of a 
static black hole event horizon. The mass at this horizon is 
$M_{\mathrm{MSH}}(r_0)= r_0/2$ and does not depend on time.

Since $\e^{2\nu} \, = \, \e^{-2\lambda} \to 1$ as $r\to \infty$, the 
geometry is asymptotically flat.

In the limit $t\to \infty$ with fixed $r$, the metric has the form 
$\e^{2\nu} \simeq \frac{t_0 \, r}{t r_0} \left(1 - \frac{r_0}{r}\right)$. 
Then, by introducing a new time coordinate $\tau$ defined by $d\tau = 
t^{-1/2} dt$ (or $\tau (t)= 2 \sqrt{t} \,$), the line element (\ref{GBiv}) 
is recast in the form \begin{eqnarray} \label{Ex3A2} ds^2 &=& - \frac{t_0 
\, r}{r_0} \left(1 - \frac{r_0}{r}\right) d\tau^2
+ \frac{r_0 \tau^2 }{4 t_0 \, r \left(1 - r_0/r \right)} \, dr^2 \nn &\, &
  + r^2 \left( d\vartheta^2 + \sin^2\vartheta \, d\varphi^2 \right)\,, 
    \end{eqnarray} in which the radial direction becomes larger like a 
    throat as the time $\tau$ (or $t$) increases. The areal radius of the 
    horizon, however, remains constant. The time-dependent factor $\left( 
    \frac{ r_0}{4 t_0 \, r \left(1 - \frac{r_0}{r}\right) }\, \tau^2 
    \right) $ multiplies only $dr^2$ and not the angular part of the line 
    element, as would happen for a central object embedded in a FLRW 
    universe \cite{Vbook} (indeed, the metric is asymptotically flat and 
    not asymptotically FLRW).

The Ricci curvature of the geometry~(\ref{Ex3A1}) is \begin{equation} 
\label{ex3a6} R = \frac{2 \left(1 + 
\frac{t_0}{t}\right)\frac{r_0}{r^3}}{\left(1
+ \frac{t r_0}{t_0 \, r}\right)^3} \left[ 1 - \frac{t}{t_0} + 2 \left( 1 - 
  \frac{t}{t_0} \right) \frac{t r_0}{t_0 \, r} + \frac{t^2 r_0^2}{t_0^2
\, r^2 }
  \right] \, , \end{equation} and is regular on the horizon $r=r_0$, 
although there are spacetime singularities at $r=0$, $t=0$, and $t\to 
\infty$.

With the ansatz~(\ref{Ex3A1}), the Hawking temperature is \begin{equation} 
\label{TH2} T=\frac{1}{4\pi r_0 \left( 1 + t / t_0 \right) } \end{equation} 
and, since the entropy $\mathcal{S}$ is \begin{equation} \label{TH3} 
\mathcal{S}=\frac{4\pi r_0^2}{4} = \frac{1}{16 \pi \left( 1 + t/t_0 
\right)^2 T^2}\, , \end{equation} the free energy and the thermodynamical 
energy are \begin{equation} \label{TH4} F= \frac{1}{16 \pi \left( 1 + t/ 
t_0 \right)^2 T}\, , \quad \quad E= \frac{1}{8 \pi \left( 1 + t/ t_0 
\right)^2 T}\, , \end{equation} respectively. Again, the black hole horizon 
is static while the Hawking temperature decreases with time (the 
adiabatic approximation can be satisfied by choosing $t_0$ and the range of 
$t$ appropriately).

We also have 
\begin{eqnarray} 
\label{TSBH6BBB} 
A&=& - \frac{\left( 1 - 
\frac{r_0}{r} \right)^2 r^2} {\left( 1 + \frac{t r_0}{t_0 \, r} \right)^2} 
\, C \nn
&=& - \frac{\left(1 - \frac{r_0}{r}\right) \left(1 +
\frac{t_0}{t}\right)t r_0} {\kappa^2 \left( 1 + \frac{t r_0}{t_0 \, 
r}\right)^2 t_0 r^3} \left[ 1 + \frac{1}{\left(1 + \frac{t r_0}{t_0 \, 
r}\right)^2}\right] \, , \nn
B&=& \frac{r_0}{\kappa^2 t_0 \, r^2 \left( 1 + \frac{t r_0}{t_0 r}\right)} 
\, ,\nn
 V&=& \frac{\left(1 + \frac{ t_0}{t}\right) r_0 }{\kappa^2 r^3} \left[
 - \frac{\frac{t}{t_0}} {\left( 1 + \frac{t r_0}{t_0 \, r}\right)^2}
+ \frac{1}{1 + \frac{t r_0}{t_0 \,r}} \right] \, . \end{eqnarray} As in the 
  case of Eqs.~(\ref{TSBH6BBB}), (\ref{TSBH7BB}), and (\ref{TSBH9BB}), also 
  this example is plagued by a phantom with negative kinetic energy because 
  $AC<0$.

\section{$f(R)$ gravity with two scalar fields} \label{sec:4} 
\setcounter{equation}{0}

The unavoidable recurrence of a phantom in Einstein's gravity with two 
scalar fields prompts us to investigate $f(R)$ gravity in the hope to 
exorcise this ghost. The equation of motion for this modified gravity is 
\begin{eqnarray} f'(R) R_{\mu\nu} - \frac{f(R)}{2}\, g_{\mu\nu} &=& 
\nabla_\mu \nabla_\nu f'(R)  - g_{\mu\nu} \Box f'(R) \nn
&\, & + \kappa^2\, T^{(\mathrm{matter)}}_{\mu\nu}\,, \label{JGRG13}
\end{eqnarray} where $T^{(\mathrm{matter)}}_{\mu\nu}$ is the matter 
energy-momentum tensor and $f'(R) \equiv df/dR$.  Assuming the 
energy-momentum tensor~(\ref{I9}) of the scalar field doublet, as in 
(\ref{TSBH2})-(\ref{TSBH5}), the components of the field 
equation~(\ref{JGRG13}) become 
\begin{widetext} 
\begin{align} 
\label{TSBH2FR0}
& - \frac{\e^{2\nu}}{2} f(R)  - \left[ - \left\{ \ddot\lambda
+ \left( \dot\lambda - \dot\nu \right) \dot\lambda \right\}
+ \e^{2\left(\nu - \lambda\right)} \left\{ \nu'' + \left(\nu' - 
  \lambda'\right)\nu' + \frac{2\nu'}{r}\right\} \right] f'(R) \nn
& + \left[ - \dot\lambda \partial_t
+ \e^{2\nu -2\lambda} \left( \partial_r^2
 + \left( - \lambda' + \frac{2}{r} \right) \partial_r \right) \right] f'(R)  
   = \, \e^{2\nu} \left( - \frac{A}{2}\e^{-2\nu} - \frac{C}{2} 
   \e^{-2\lambda} - V \right) \, ,\\
& \frac{\e^{2\lambda}}{2} f(R)
 - \left[ \e^{-2\left( \nu - \lambda \right)} \left\{ \ddot\lambda
+ \left( \dot\lambda - \dot\nu \right) \dot\lambda \right\}
 - \left\{ \nu'' + \left(\nu' - \lambda'\right)\nu' \right\}
+ \frac{2 \lambda'}{r} \right] f'(R) \nn 
& - \left[ - \e^{2\lambda -2\nu} \left( \partial_t^2 -\dot\nu 
   \partial_t\right)
+  \left( \nu' + \frac{2}{r} \right) \partial_r \right] f'(R) =
 - \e^{2\lambda} \left( \frac{A}{2}\e^{-2\nu} + \frac{C}{2} \e^{-2\lambda}
 - V \right) \, ,\\ 
\label{TSBH4FR} & \frac{1}{2} f(R)
 - \frac{1}{r^2} \left\{ 1 + \left\{ - 1 - r \left(\nu'
 - \lambda' \right)\right\}\e^{-2\lambda}\right\} f'(R) \nn 
& - \left[ - \e^{-2\nu} \left( \partial_t^2
+ \left( -\dot\nu + \dot\lambda \right)\partial_t\right)
+ \e^{-2\lambda} \left( \partial_r^2
+ \left( \nu' - \lambda' + \frac{1}{r} \right)\partial_r \right) \right] 
  f'(R) \nn
& = - \left( \frac{A}{2}\e^{-2\nu} - \frac{C}{2} \e^{-2\lambda} - V
\right) \, ,\\
& - \frac{2\dot\lambda}{r} f'(R) + \left( { \partial_t \partial_r}
 - \nu'\partial_t - \dot\lambda \partial_r \right) f'(R) = { -} B \,.  
\end{align} 
Then, we have 
\begin{align} 
\label{TSBHFR1} 
A=& \left[ -    \left\{ \ddot\lambda
+ \left( \dot\lambda - \dot\nu \right) \dot\lambda \right\}
+ \e^{2\left(\nu - \lambda\right)} \left\{ \nu'' + \left(\nu' - 
  \lambda'\right)\nu' + \frac{2\nu'}{r}\right\} \right. \nn
& \left. - \frac{\e^{2\nu}}{r^2} \left\{ 1 + \left\{ - 1 - r \left(\nu'
 - \lambda' \right)\right\}\e^{-2\lambda}\right\} \right] f'(R) \nn 
& - \left[ - \partial_t^2 + \dot\nu \partial_t
+ \e^{2\nu -2\lambda} \left( \nu' + \frac{1}{r} \right)\partial_r \right] 
  f'(R) \, ,\\  
\label{TSBHFR4} B=& \frac{2\dot\lambda}{r} f'(R) - \left( \partial_t 
\partial_r - \nu'\partial_t - \dot\lambda \partial_r \right) f'(R)\, ,\\
\label{TSBHFR2} C=& \left[ \e^{-2\left( \nu - \lambda \right)} \left[ 
\ddot\lambda
+ \left( \dot\lambda - \dot\nu \right) \dot\lambda \right]
- \left[ \nu'' + \left(\nu' - \lambda'\right)\nu' \right]
+ \frac{2 \lambda'}{r} \right. \nn 
& \left. + 
  \frac{\e^{2\lambda}}{r^2} \left\{ 1 + \left\{ - 1 - r \left(\nu' - 
  \lambda' \right)\right\}\e^{-2\lambda}\right\} \right] f'(R) \nn
& + \left[ - \e^{-2\nu + 2\lambda} \dot\lambda \partial_t
+ \left( \partial_r^2
+ \left( - \lambda' - \frac{1}{r} \right)\partial_r \right) \right] f'(R)  
  \, ,\\ 
\label{TSBHFR3} V=& \, \frac{f(R)}{2} { +} \left[ - \e^{-2\nu} \left\{ 
\ddot\lambda
+ \left( \dot\lambda - \dot\nu \right) \dot\lambda \right\}
+ \e^{- 2 \lambda} \left\{ \nu'' + \left(\nu' - \lambda'\right)\nu' + 
  \frac{ \nu' - \lambda'}{r}\right\} \right] f'(R) \nn
& +  \frac{1}{2} \left[ \e^{-2\nu} \left\{ \partial_t^2
 - \left(\dot\nu - \dot\lambda\right) \partial_t \right\}
 - \e^{-2\lambda} \left\{ \partial_r^2 + \left( \nu' - \lambda' \frac{4}{r} 
   \right)\partial_r \right\} \right] f'(R) \, . 
\end{align} 
\end{widetext}
Again, if we replace $t$ and $r$ with $\phi$ and $\chi$ in the right hand 
sides of Eqs.~(\ref{TSBHFR1})-(\ref{TSBHFR3}), we obtain $A$, $B$, $C$, and 
$V$ as functions of $\phi$ and $\chi$. Conversely if we assign $A$, $B$, 
$C$, and $V$, the model has a solution in the form of the spherically 
symmetric configuration~(\ref{GBiv}) corresponding to arbitrary $\nu$ and 
$\lambda$.

In the case $\lambda=-\nu$ (or $g_{tt}g_{rr}=-1$) already discussed, we 
find 
\begin{widetext} 
\begin{align} 
\label{TSBHFR1BBB} 
A=& \left[ - 
\frac{1}{2} \, \e^{2\nu} \, \frac{d^2 \left( \e^{-2\nu} \right) }{dt^2} + 
\e^{2\nu} \left\{ \frac{1}{2} \frac{d^2 \left( \e^{2\nu} \right) }{dr^2} + 
\frac{1}{r} \, \frac{d \left( \e^{2\nu} \right) }{dr} \right\} - 
\frac{\e^{2\nu}}{r^2} \left\{ 1 + \left( - \e^{2\nu} - r \, \frac{d \left( 
\e^{2\nu} \right) }{dr} \right) \right\} \right] f'(R) \nn
& - \left[ - \partial_t^2 - \frac{ \e^{2\nu} }{2} \, \frac{d 
   \left(\e^{-2\nu}\right) }{dt} \, \partial_t + \left( \e^{2\nu} \, 
   \frac{d \left( \e^{2\nu} \right)}{dr}
 + \frac{\e^{4\nu}}{r} \right) \partial_r \right] f'(R) \, ,\\
\label{TSBHFR4}
B=& \, \frac{\e^{2\nu}}{r} \, \frac{d \left( \e^{-2\nu}\right)}{dt} \, 
f'(R) - \left( { \partial_t \partial_r } - \frac{\e^{-2\nu}}{2} \frac{d 
\left( \e^{2\nu} \right) }{dr} \partial_t - \frac{\e^{2\nu}}{2} \, \frac{d 
\left(\e^{-2\nu}\right)}{dt} \, \partial_r \right) f'(R) \, ,\\
C=& \left[ \frac{ \e^{-2\nu} }{2} \, \frac{d^2 \left( 
\e^{-2\nu}\right)}{dt^2} - \frac{\e^{-2\nu}}{2} \, \frac{d^2 \left( 
\e^{2\nu}\right)}{dr^2}
- \frac{\e^{-2\nu}}{2} \, \frac{d \left( \e^{2\nu}\right) }{dr}
+ \frac{\e^{-2\nu}}{r^2} \left[ 1 + \left\{ - 1 - r \e^{-2\nu}\, \frac{d 
  \left( \e^{2\nu}\right) }{dr} \right] \e^{2\nu}\right\} \right] f'(R) \nn
& + \left[ - \frac{\e^{-2\nu}}{2}\, \frac{d \left(\e^{-2\nu}\right)}{dt}
\, \partial_t + \left( \partial_r^2
+ \left( \frac{\e^{-2\nu}}{2} \, \frac{d \left( \e^{2\nu}\right) }{dr} - 
  \frac{1}{r} \right) \partial_r \right) \right] f'(R) \, ,\\
V=& \, \frac{f(R)}{2} + \left[ \frac{1}{2} \, \frac{d^2 \left( 
\e^{-2\nu}\right) }{dt^2}
+ \frac{1}{2}\, \frac{d^2 \left( \e^{2\nu}\right) }{dr^2}
+ \frac{1}{r}\, \frac{d \left( \e^{2\nu}\right) }{dr} \right] f'(R) \nn
& + \frac{1}{2} \left[ \e^{-2\nu} \partial_t^2
 + \frac{d \left( \e^{-2\nu}\right)}{dt}
\partial_t
 -  \e^{2\nu} \partial_r^2 - \left( \frac{d \left( \e^{2\nu}\right) }{dr}
+ \frac{4}{r} \right) \partial_r \right] f'(R) \, . 
\end{align} 
  \end{widetext} Contrary to the Einstein gravity case~(\ref{TSBH6BB}), 
  (\ref{TSBH7BB}), and~(\ref{TSBH9BB}), it is now possible to avoid the 
  ghost with a suitable choice of the Lagrangian density $f(R)$.

\subsection{Example~1}

With our previous ansatz~(\ref{E1}) (with $\dot{M}\ll1$ to ensure 
the adiabatic approximation), Eqs.~(\ref{TSBHFR1BBB})-(\ref{TSBHFR4}) 
acquire the form 
\begin{widetext} 
\begin{align} 
\label{TSBHFR5} 
A=& { -} \left[ \frac{\ddot M}{2r} + \frac{\left( \frac{\dot M}{r} \right)^2} 
{2\left( 1 - \frac{M}{r} \right)} \right] f'(R) \nn
& - \left[ - \partial_t^2
 - \frac{\frac{\dot M}{r}}{2 \left( 1 - \frac{M}{r} \right)} \partial_t
 - \frac{1}{r} \left( 1 - \frac{2M}{r} \right)  \partial_r \right] f'(R) \, 
   ,\\
\label{TSBHFR6} C=& \, \left( 1 - \frac{M}{r} \right)^{-2}\left[ 
\frac{\ddot M}{2r} + \frac{\left( \frac{\dot M}{r} \right)^2}{2\left( 1 - 
\frac{M}{r} \right)} \right] f'(R) \nn
& { -} \left[ - \left( 1 - \frac{M}{r} \right)^{-3} \frac{\dot M}{2r}
\partial_t + \left( \partial_r^2 - \frac{1}{2r} \left( 1 - \frac{M}{r} 
\right)^{-1} \left( 1 - \frac{3M}{r} \right) \partial_r \right) \right] 
f'(R) \, ,\\
\label{TSBHFR7} 
V=& \, \frac{f(R)}{2} + \left( 1 - \frac{M}{r} \right)^{-1} 
\left( \frac{\ddot M}{2r} + \frac{\left( \frac{\dot M}{r} \right)^2} 
{2\left( 1 - \frac{M(t)}{r} \right)} \right) f'(R) \nn
&  + \frac{1}{2} \left[ \left( 1 - \frac{M}{r} \right)^{-1} \left\{ 
    \partial_t^2
 + \frac{\frac{\dot M}{r}}{1 - \frac{M}{r} } \partial_t \right\}
 - \left( 1 - \frac{M}{r} \right) \left\{ \partial_r^2
+ \frac{\frac{M}{r^2} + \frac{4}{r}} {1 - \frac{M}{r}} \partial_r \right\} 
  \right] f'(R)  \, ,\\
\label{TSBHFR8} B=& \, \frac{1}{\kappa^2} \left( 1 - \frac{M}{r} 
\right)^{-1}\left( \frac{M'}{r} - \partial_t \partial_r
+ \frac{M}{2r^2} \, \partial_t \,
+ \frac{\dot M}{2r} \, \partial_r \right) f'(R)\, . 
\end{align} 
\end{widetext} 
Even in $f(R)$ gravity, it is difficult to avoid the ghost and we need to 
specify the model explicitly.

As done for Eqs.~(\ref{T1})-(\ref{T4}), let us discuss the thermodynamics 
of the spacetime~(\ref{E1}). Assuming again a quasi-static evolution  
($\dot{M} \ll 1$) and treating the time $t$ as constant, the temperature 
is 
\begin{equation} 
\label{T1FR} T=\frac{1}{8 \pi M(t)}\,, 
\end{equation} 
whereas the Bekenstein-Hawking entropy reads 
\begin{equation} \label{T2FR} 
\mathcal{S} = \frac{4\pi \left[ 2 M(t) \right]^2 f'\left( R \left( r \to 2M 
\left(t \right) \right) \right)} {4} = \frac{f'\left( +\infty 
\right)}{16\pi T^2}\, . 
\end{equation} 
In order to obtain a non-trivial 
entropy, $f'\left( +\infty \right)$ must be finite. The free energy is 
\begin{equation} \label{T3FR} 
F= \frac{f'\left( +\infty \right)}{16\pi T} 
\,, 
\end{equation} 
where we have chosen the integration constant so that 
$F$ vanishes in the limit $T\to \infty$ (or $M(t)\to 0$). The thermal 
energy is 
\begin{equation} \label{T4FR} 
E = \frac{f'\left( +\infty 
\right)}{8\pi T} =f'\left( +\infty \right)M(t) \,, 
\end{equation} 
where a 
correction $f'\left( +\infty \right)$ appears, as in the expression of $T$. 
This correction must be present, as one realizes by remembering that $f(R)$ 
gravity is equivalent to an $\omega=0$ Brans-Dicke theory 
\cite{Brans:1961sx} with the Brans-Dicke scalar $\phi_\mathrm{BD}=f'(R)$ 
\cite{Sotiriou:2008rp,Nojiri:2017ncd, Nojiri:2006ri}, which plays the role 
of the inverse of the effective gravitational coupling \cite{Brans:1961sx}, 
$\phi_\mathrm{BD}=G_\mathrm{eff}^{-1}$, and with a potential 
$V(\phi_\mathrm{BD})$. Restoring Newton's constant $G$, the corresponding 
expressions in Brans-Dicke gravity contain $8\pi G\mathrm{eff} =8\pi 
/f'(R)$.

\subsection{Example 2}

Moving on to the ansatz~(\ref{Ex2a}), we find 
\begin{widetext} 
\begin{align} 
\label{TSBHFR1EX2} 
A=& \, - \frac{t_0}{t r^2} \left( 1 - 
\frac{r_0}{r} \right) \left( 1 - \frac{t_0}{t} \right) f'(R) 
+ \left[ - \partial_t^2 - \frac{1}{2t} \, \partial_t
+ \left( 1 - \frac{r_0}{r} \right) \frac{t_0}{t r} \partial_r \right] f'(R)  
  \,,\\
\label{TSBHFR2EX2} C=& \, \left( 1 - \frac{r_0}{r} \right)^{-1} 
\frac{1}{r^2} \left( \frac{t}{t_0} - 1 \right) f'(R) \nn
& - \left[ - \left( 1 - \frac{r_0}{r} \right)^{-2} \frac{t}{2t_0^2} \, 
   \partial_t + \left( \partial_r^2 + \frac{1}{2r} \left( 1 - 
   \frac{r_0}{r} \right)^{-1} \left( -2 + \frac{3r_0}{r} \right)  
   \partial_r \right) \right] f'(R) \, ,\\
\label{TSBHFR3EX2} V=& \, \frac{f(R)}{2}
+ \frac{1}{2} \left[ \left( 1 - \frac{r_0}{r} \right)^{-1} \frac{t}{t_0} 
  \left\{ \partial_t^2 + \frac{1}{t} \partial_t \right\}
+ \left( 1 - \frac{r_0}{r} \right) \frac{t_0}{t}\left\{ \partial_r^2
 - \frac{\frac{r_0}{r^2} + \frac{4}{r}} {1 - \frac{r_0}{r}} \partial_r 
   \right\} \right] f'(R) \, ,\\
\label{TSBHFR4EX2} B=& \, { \frac{ f'(R)}{tr} -} \left( { \partial_t 
\partial_r}
 - \frac{\frac{r_0}{r^2}}{2\left( 1 - \frac{r_0}{r} \right)} \, \partial_t
 - \frac{1}{2t}\, \partial_r \right) f'(R)\, . 
\end{align} 
The Ricci 
   scalar~(\ref{Ex2c}) yields \begin{align} \label{TSBHFR1EX2B} A=& \, - 
   \frac{t_0}{t r^2} \left( 1 - \frac{r_0}{r} \right) \left( 1 - 
   \frac{t_0}{t} \right) f'(R)
+ \frac{4t_0^2}{r^4 t^4} \, f'''(R) \nn 
& + \left[ \frac{3t_0}{r^2 t^3} - 4 \left( 1 - \frac{r_0}{r} \right)  
   \frac{t_0^2}{t^2 r^4} \left( 1 - \frac{t_0}{t} \right) \right] f''(R)  
   \, ,\\
C=& \, \frac{t}{t_0 \,r^2} \left( 1 - \frac{r_0}{r} \right)^{-1} \left( 1
 - \frac{t_0}{t} \right) f'(R)
+ \frac{16}{r^6} \left( 1 - \frac{t_0}{t} \right)^2 f'''(R) \nn 
& - \left[ - \left( 1 - \frac{r_0}{r} \right)^{-2} \frac{1}{t_0 \,r^2 t}
+ \frac{2}{r^4} \left( 1 - \frac{t_0}{t} \right) \left( 1 - \frac{r_0}{r} 
  \right)^{-1} \left( 8 - \frac{9r_0}{r} \right) \right] f''(R) \,,\\
\label{TSBHFR3EX2} V=& \, \frac{f(R)}{2} + \frac{ 2 t_0}{r^4 t^3} \left[ 
\left( 1 - \frac{r_0}{r} \right)^{-1} + \frac{ 4 t_0}{r^6 t} \left( 1 - 
\frac{r_0}{r} \right) \left( 1 - \frac{t_0}{t} \right)^2 \right] f'''(R) 
\nn
& - \left[ - \frac{ 2}{r^2 t^2}\left( 1 - \frac{r_0}{r} \right)^{-1}
+  \frac{t_0}{r^4 t} \left( 1 - \frac{t_0}{t} \right) \left( 28 - 
  \frac{8r_0}{r} \right) \right] f''(R) \, ,\\
\label{TSBHFR4EX2} B=& \, \frac{f'(R)}{tr} - \left[ - \frac{4t_0}{r^3 t^2}
 - \frac{r_0 t_0}{r^4 t^2} \left( 1 - \frac{r_0}{r} \right)^{-1}
+ \frac{2}{r^3 t} \left( 1 - \frac{t_0}{t} \right) \right] f''(R) \nn
& + \frac{8t_0}{r^3 t^2} \left( 1 - \frac{r_0}{r} \right) f'''(R) \, .  
\end{align} 
\end{widetext} 
In order to check whether the ghost can be avoided, let us examine the 
functions $A$, $B$, and $C$ in the regime of large $r$, in which one finds 
\begin{align} 
\label{TSBHFR1EX2Bas} 
A\sim & \, { -} \frac{t_0}{t r^2} \left( 1 - \frac{t_0}{t} \right) f'(R)
+ \frac{3t_0}{r^2 t^3} \, f''(R) { -} \frac{4t_0^2}{r^4 t^4}\, f'''(R) 
\,,\\
\label{TSBHFR4EX2as} 
B\sim & \, \frac{f'(R)}{tr} - \frac{2}{r^3 t} \left( 1
- \frac{t_0}{t} \right) f''(R) + \frac{8t_0}{r^3 t^2} \, f'''(R) \, ,\\
\label{TSBHFR2EX2as} C\sim & \, \frac{t}{t_0 \, r^2} \left( 1 - 
\frac{t_0}{t} \right) f'(R) + \frac{1}{t_0 \, r^2 t} \, f''(R)\nn
& - \frac{16}{r^6} \left( 1 - \frac{t_0}{t} \right)^2 f'''(R) \,.  
\end{align} 
The popular power-law choice (\cite{Nojiri:2010wj,Nzioki:2010nj,  
Park:2010da,Nojiri:2007cq,Nojiri:2009kx,Odintsov:2020thl,  
Odintsov:2020nwm,Nojiri:2017qvx,  
Capozziello:2009jg,Dunsby:2009zz,Faraoni:2009xb,  
Goheer:2009ss,Goheer:2008tn,Ananda:2008tx,Carloni:2007br,
Carloni:2008jy,Ananda:2008gs,Goheer:2007wx,Ananda:2007xh, 
Carloni:2007yv,Amendola:2006we,Goheer:2007wu,Clifton:2007ih,
Corda:2007hi,Leach:2007ss,Carloni:2006mr,Clifton:2006kc, 
Capozziello:2006dp,Leach:2006br,Clifton:2006ug,Carloni:2004kp, 
Clifton:2005aj,Furey:2004rq,Pavlov:1997xf,Ferraris:1992dx, 
Burlankov1989,Leon:2010pu,Bisabr:2010sq,Carloni:2010tv, 
Capozziello:2007vd,Corda:2008uh,Capozziello:2008rq, 
Cervantes:2009tb,Martins:2007uf,Capozziello:2006ph, 
Mendoza:2006hs,Saffari2007,Corda:2007tz, 
Capozziello:2003tk,Capozziello:2002rd,Faraoni:2011pm}, see 
\cite{Barrow:2005dn,Zakharov:2006uq} for Solar System constraints on the 
parameter $n$) \begin{equation} \label{Ex2Fn1} f'(R) \sim f_0 R^n = f_0 
\left[ \frac{2}{r^2} \left( 1 - \frac{t_0}{t} \right) \right]^n \, , 
\end{equation} produces \begin{align} \label{TSBHFR1EX2Bas2} A\sim & { -} 
\frac{ f_0 n t_0}{r^4 t^3} \left[ \frac{2}{r^2} \left( 1 - \frac{t_0}{t} 
\right) \right]^{n-2} \left[ - 6 + \left( 2 + 4n \right) \frac{t_0}{t} 
\right] \, , \\
\label{TSBHFR2EX2as2} C\sim & \, \frac{f_0 n}{t_0 \, r^2 t} \left[ 
\frac{2}{r^2} \left( 1 - \frac{t_0}{t} \right)  \right]^{n-1} \, ,\\
\label{TSBHFR4EX2as2} B\sim & \, \frac{f_0 }{tr} \left[ \frac{2}{r^2} 
\left( 1 - \frac{t_0}{t} \right) \right]^n - \frac{2 f_0 n}{r^3 t} \left( 1
- \frac{t_0}{t} \right) \left[ \frac{2}{r^2} \left( 1 - \frac{t_0}{t} 
  \right) \right]^{n-1}\, . \end{align} The special case $n=1$ gives 
  \begin{equation} \label{TSBHFR1EX2Bas3} A\sim -\frac{3 f_0 t_0}{r^4 t^3} 
  \, , \quad C\sim - \frac{f_1}{t_0 \,r^2 t} \, , \quad B\sim \mathcal{O} 
  \left( r^{-4} \right) \,; \end{equation} therefore, if $f_0<0$, both $A$ 
  and $B$ are positive and \begin{equation} \label{TSBHFR1EX2Bas4} AC - B^2 
  \sim \frac{3f_0^2}{r^6 t^4} \, , \end{equation} which tells us that, at 
  least as long as $r$ is sufficiently large, there is no ghost. Thus, it 
  seems possible to avoid ghosts in $f(R)$ gravity.

Again, we obtain the thermodynamics of this spacetime in the adiabatic 
approximation (assuming $|t| \gg t_0$). The temperature and the 
Bekenstein-Hawking entropy are \begin{equation} \label{T1e2FR} 
T=\frac{t_0}{4 \pi t r_0}\, , \end{equation} \begin{equation} \label{T2e2B} 
\mathcal{S} = \frac{4\pi f' \left( R \left( r \to r_0 \right) \right) 
r_0^2}{4} = \pi f' \left( \frac{2}{r_0^2} \left( 1 - \frac{t_0}{t} \right) 
\right) r_0^2 \, , \end{equation} and they are both time-dependent. 
Eliminating $r_0$ with Eq.~(\ref{T1e2}) yields \begin{equation} 
\label{T2Be2B} \mathcal{S} = \frac{t_0^2}{4 t^2 T^2} f' \left( 
\frac{32\pi^2 t^2 T^2}{t_0^2} \left( 1 - \frac{t_0}{t} \right) \right) \, . 
\end{equation} For this explicit form of $f(R)$, the use of $dF/dT=- 
\mathcal{S}$ and $E=F+TS$ produces the non-trivial expressions of the free 
energy and the thermodynamical energy \begin{eqnarray} \label{T3FRex2} F
&=& - \int \mathcal{S} dT = - \int \frac{t_0^2 f' \left( \frac{32\pi^2 t^2
T^2}{t_0^2} \left( 1 - \frac{t_0}{t} \right) \right)}{4 t^2 T^2} \, dT \, , 
\nn
E &=& F+T\mathcal{S} = \frac{t_0^2f' \left( \frac{32\pi^2 t^2 T^2}{t_0^2} 
\left( 1 - \frac{t_0}{t} \right) \right)}{4 t^2 T} \nn
&\, & - \int \frac{t_0^2f'
\left( \frac{32\pi^2 t^2 T^2}{t_0^2} \left( 1 - \frac{t_0}{t} \right) 
\right)}{4 t^2 T^2} \, dT \, . \end{eqnarray} By construction, these 
quantities satisfy the first law of thermodynamics $dE=Td\mathcal{S}=dQ$.  
For the model~(\ref{Ex2Fn1}), we find \begin{eqnarray} \label{T4FRex2} F
&=& - \frac{ t_0^2 f_0 \left( \frac{32\pi^2 t^2 T^2}{t_0^2} \left( 1 -
\frac{t_0}{t} \right) \right)^n}{4 \left( 2n - 1 \right) t^2 T} \, , \nn 
E&=& \frac{ \left( n - 1 \right) t_0^2 f_0 \left( \frac{32\pi^2 t^2 
T^2}{t_0^2} \left( 1 - \frac{t_0}{t} \right)\right)^n}{2 \left( 2n - 1 
\right) t^2 T} \, . \end{eqnarray} When $t>t_0$, the thermodynamical energy 
$E$ is positive if $f_0>0$ and $n>1$ or $n< 1/2$, or if $f_0<0$ and $ 1/2 
<n<1$. Moreover, when $t<t_0$ and $n$ is an integer, $E$ is positive in the 
following cases:

\begin{itemize}

\item $f_0>0$, $n$ is an even integer, and $n\geq 2$ or $n\leq 0$; \item 
$f_0<0$ and $n=0$ (which corresponds to Einstein's gravity);

\item $f_0<0$, $n$ is an odd integer, and $n\geq 3$ or $n\leq -1$.

\end{itemize}

The cases with negative $E$ may be unphysical.

\subsection{Example 3}

For the example~(\ref{Ex3A1}), we find 
\begin{widetext} 
\begin{align} 
\label{TSBHFR1BB} 
A=& \, \frac{\left(1 - \frac{r_0}{r}\right)\left(1 + 
\frac{t_0}{t}\right) \left( \frac{t}{t_0} - 1 + \frac{t r_0}{t_0 r} 
\right)\frac{t r_0^2}{t_0 r^4}} {\left(1 + \frac{t r_0}{t_0 r}\right)^4} \, 
f'(R) 
 - \left[ - \partial_t^2 - \frac{\frac{r_0}{2t_0 r}}{ 1 + \frac{t r_0}{t_0 
   r}} \, \partial_t
+ \frac{\left(1 - \frac{r_0}{r}\right)  \left( 1 + \frac{t r_0}{t_0 r} 
  \right)
}{r\left(1 +
  \frac{t r_0}{t_0 r}\right)^3} \partial_r \right] f'(R) \, , \\
\label{TSBHFR4BB} B=& \, \frac{\frac{r_0}{t_0 r^2}}{ 1 + \frac{t r_0}{t_0 
r}} \, f'(R)
- \left( \partial_t \partial_r
- \frac{\left(1 + \frac{t_0}{t}\right)\frac{t r_0}{2 t_0 r^2}} {\left( 1 - 
  \frac{r_0}{r} \right)\left(1 + \frac{t r_0}{t_0 r}\right)} \partial_t
- \frac{\frac{r_0}{2t_0 r}}{ 1 + \frac{t r_0}{t_0 r}} \partial_r \right)
   f'(R) \, , \\
\label{TSBHFR2BB} C=& - \frac{\left(1 + \frac{t_0}{t}\right) \left( 
\frac{t}{t_0} - 1 + \frac{t r_0}{t_0 r} \right)\frac{t r_0^2}{t_0 r^4}} 
{\left(1 - \frac{r_0}{r}\right) \left(1 + \frac{t r_0}{t_0 r}\right)^2} \, 
f'(R) \nn
& + \left[ - \frac{ \left(1 + \frac{t r_0}{t_0 r}\right) \frac{r_0}{2 t_0 
   r}} {\left(1 - \frac{r_0}{r}\right)^2} \, \partial_t
+ \left( \partial_r^2
+ \left( \frac{\left(1 + \frac{t_0}{t}\right) \frac{t r_0}{t_0 r^2}} 
  {\left( 1 - \frac{r_0}{r} \right) \left(1 + \frac{t r_0}{t_0 r}\right)}
- \frac{1}{r} \right)\partial_r \right) \right] f'(R) \, ,\\ 
  \label{TSBHFR3BB} V=& \, \frac{f(R)f'(R)}{2} \left[ - \frac{\left(1 + 
  \frac{t_0}{t}\right) \frac{t r_0}{t_0 r^3}} {\left(1 + \frac{t r_0}{t_0 
  r}\right)^3}
+ \frac{\left(1 + \frac{t_0}{t}\right)\frac{t r_0}{t_0 r^3}} { \left(1 + 
  \frac{t r_0}{t_0 r}\right)^2} \right] \nn
& +  \frac{1}{2} \left[ \frac{ 1 + \frac{t r_0}{t_0 r}}{1 - \frac{r_0}{r}} 
   \partial_t^2
 - \frac{ \frac{r_0}{t_0 r}}{1 - \frac{r_0}{r}} \, \partial_t
+ \frac{1 - \frac{r_0}{r}}{ 1 + \frac{t r_0}{t_0 r}} \, \partial_r^2 
  -\left( \frac{\left(1 + \frac{t_0}{t}\right)\frac{t r_0}{t_0 r^2}}
{ \left(1 + \frac{t r_0}{t_0 r}\right)^2} + \frac{4}{r} \right)  
\partial_r \right] f'(R) \,. 
\end{align} 
\end{widetext} 
The Hawking 
temperature and the entropy now read, in the adiabatic approximation 
$|t| \gg t_0$) 
\begin{equation} 
\label{TH2FR} 
T=\frac{f'\left( R \left( 
r\to r_0, t \right) \right)}{4\pi r_0 \left( 1
+ \frac{t}{t_0}\right) } = \frac{f'\left( \frac{\frac{2}{r_0^2}\left(1 - 
  \frac{t}{t_0} - \frac{t^2}{t_0^2} \right)} {\left(1 + 
  \frac{t}{t_0}\right)^2} \right)}{4\pi r_0 \left( 1 + \frac{t}{t_0}\right)
  } \, , 
\end{equation} 
\begin{equation} \label{TH3B} 
    \mathcal{S}=\frac{4\pi r_0^2}{4} = \frac{f'\left( 
    \frac{\frac{2}{r_0^2}\left(1 - \frac{t}{t_0} - \frac{t^2}{t_0^2} 
    \right)} {\left(1 + \frac{t}{t_0}\right)^2} \right)^2}{16 \pi \left( 1 
    + \frac{t}{t_0}\right)^2 T^2}\, . 
\end{equation} 
Non-trivial expressions of the free energy $F$ and the thermodynamical 
energy $E$ can be obtained as done with Eqs.~(\ref{T3FRex2}) and 
(\ref{T4FRex2}).  See Appendix~\ref{AppendixB} for further comments 
on $f(R)$ gravity with two scalar fields.

\section{Spherically symmetric, time-dependent geometry in pure $f(R)$ 
gravity} \label{sec:5} \setcounter{equation}{0}

It is difficult to avoid the ghost degree of freedom even in pure $f(R)$ 
gravity. In this section, in the framework of pure $f(R)$ gravity, we 
construct a model which admits a spherically symmetric and time-dependent 
solution.

Although the sought-for solution is time-dependent, we begin with a model 
that gives a static solution. We choose $A=\alpha^2$, $B=C=0$, and 
$V=V_0=$~constant; then Eqs.~(\ref{TSBH2})-(\ref{TSBH5}) assume the form 
\begin{widetext} \begin{align} \label{TSBH2FR}
& \frac{\e^{2\left(\nu -
\lambda\right)}}{\kappa^2} \left( \frac{2\lambda'}{r} + \frac{\e^{2\lambda}
- 1}{r^2} \right) = - \e^{2\nu} \left( \frac{\alpha^2}{2}\, \e^{-2\nu} + 
  V_0 \right) \, , \\
\label{TSBH3FR}
& \frac{1}{\kappa^2} \left( \frac{2\nu'}{r} -
\frac{\e^{2\lambda} - 1}{r^2} \right) = - \e^{2\lambda} \left( 
\frac{\alpha^2}{2}\e^{-2\nu} - V_0 \right) \, , \\
\label{TSBH4FR}
&\frac{1}{\kappa^2} \left[ \e^{-2\lambda}\left( r
\left(\nu' - \lambda' \right) + r^2 \nu'' + r^2 \left( \nu' - \lambda' 
\right) \nu' \right) \right] = - r^2 \left( \frac{\alpha^2}{2}\e^{-2\nu}
- V_0 \right) \, . \end{align} \end{widetext} Equations~(\ref{TSBH2FR}) and 
  (\ref{TSBH3FR}) give \begin{align} \label{FRR4} \lambda' =& - 
  \frac{\left( \e^{2\lambda} - 1\right)}{2r}
 - \frac{\kappa^2 r \e^{2\lambda}}{2} \left( \frac{\alpha^2 \, \e^{-2\nu} 
   }{2}
+ V_0 \right) \, , \\ 
\label{FRR5} \nu' =& \, \frac{\e^{2\lambda} - 
  1}{2r} - \frac{\kappa^2 r \e^{2\lambda}}{2} \left( 
  \frac{\alpha^2}{2}\e^{-2\nu} - V_0 \right) \, . 
\end{align} 
By using Eqs.~(\ref{FRR4}) and~(\ref{FRR5}), one checks that 
Eq.~(\ref{TSBH4FR}) is automatically satisfied. Therefore, we can 
choose~(\ref{FRR4} and~(\ref{FRR5}) as our two independent equations. By 
solving them with physically realistic boundary conditions, one obtains a 
spherically symmetric and static solution $\nu=\nu_\mathrm{ss}(r)$ and 
$\lambda=\lambda_\mathrm{ss}(r)$.

Now the action has the form 
\begin{equation} 
\label{I8BB} 
S_{\phi} = \int d^4 x \sqrt{-g} \left( \frac{R}{2\kappa^2}
 - \frac{\alpha^2}{2} \, \partial_\mu \phi \partial^\mu \phi - V_0 \right)  
   \, 
\end{equation} 
describing GR with a single, canonical, scalar field. Let us examine the 
relation between the model~(\ref{I8BB}) and $f(R)$ gravity. Under the 
conformal transformation $g_{\mu\nu}\to \tilde{g}_{\mu\nu}= \e^\rho 
g_{\mu\nu}$, the Ricci scalar transforms according to \cite{Wald:1984rg} 
\begin{equation} 
\label{E4} 
R \to  \tilde{R}= \left(R - 3\Box \rho - \frac{3}{2} \, \partial^\mu \rho  
\partial_\mu \rho \right)\e^{-\rho}  
\end{equation} 
and the action~(\ref{I8BB}) is mapped into 
\begin{eqnarray} 
S_{\phi} &=& \int d^4 x \sqrt{-g} \, \e^{2\rho} \left[ \frac{1}{2\kappa^2} 
\left( R - 3\Box \rho - \frac{3}{2}\partial^\mu \rho \partial_\mu \rho 
   \right)\e^{-\rho} \right. \nn
&\, & \left. - \frac{\alpha^2}{2} \e^{-\rho} \partial_\mu \phi
\partial^\mu \phi
- V_0 \right] \,. \label{SII6phi} 
\end{eqnarray} 
The choice $\rho = \phi/\sqrt{3}$ yields 
\begin{equation} 
\label{SII7phi} 
S_{\phi} = \int  d^4 x \sqrt{-g} \left( \frac{1}{2\kappa^2}\, 
\e^{\frac{\phi}{\sqrt{3}}} R - \e^{\frac{2\phi}{\sqrt{3}}} \tilde V_0 
\right) 
\end{equation} 
and the variation of the action~(\ref{SII7phi}) with respect to $\phi$ 
yields the algebraic equation 
\begin{equation} \label{SII8phi} 
\frac{1}{2\kappa^2}  \, \e^{\frac{\phi}{\sqrt{3}}} R - 2 \, 
\e^{\frac{2\phi}{\sqrt{3}}} \, V_0 =0 
\end{equation} 
for  $\e^{\phi/\sqrt{3}}$, which has the unique root 
\begin{equation}  \label{FRRR1} 
\e^{ \phi/\sqrt{3} } = \frac{R}{4\kappa^2 V_0} \, .  
\end{equation} 
By substituting Eq.~(\ref{FRRR1}) into the action~(\ref{SII7phi}), one 
obtains the $R^2$ gravity model 
\begin{equation} 
\label{SII7phiRsq} 
S_{\phi} = - \frac{1}{8\kappa^4 V_0}  \int d^4 x \sqrt{-g} \, R^2\, . 
\end{equation} 
Due to the conformal transformation $g_{\mu\nu}\to \tilde{g}_{\mu\nu} 
\equiv \e^\rho \,  g_{\mu\nu}=\e^{\phi/\sqrt{3}} \, g_{\mu\nu}$, the 
spacetime metric that solves Eqs.~(\ref{FRR4}) and~(\ref{FRR5}) differs 
from $g_{\mu\nu}$ by the conformal factor $\e^{- \phi/\sqrt{3} 
}=\e^{-t/\sqrt{3}}$, or 
\begin{eqnarray} 
\label{GBivB} 
ds^2 &=& \e^{-\frac{t}{\sqrt{3}}} \left[ - \e^{2\nu_\mathrm{ss} (r)} dt^2
+ \e^{2\lambda_\mathrm{ss} (r)} dr^2 \right.\nn &\, & \left. + r^2 \left( 
  d\vartheta^2 + \sin^2\vartheta \, d\varphi^2
\right)\right] \, , 
\end{eqnarray} 
a time-dependent line element.

Let us discuss now the solutions~(\ref{FRR4}) and~(\ref{FRR5}) in more 
detail. Define $X_\pm= \e^{\nu \pm \lambda}$ and rewrite these equations as 
\begin{eqnarray} \label{FRR4C} X_+'&=& { -} \kappa^2 r X_+^2 X_-^{-1} 
\left(\frac{\alpha^2}{2} X_+^{-1} X_-^{-1} + V_0\right) \, , \nn
 X_-' &=& \frac{X_- \left( X_+ X_-^{-1} - 1\right)}{2r} \, . 
\end{eqnarray} 
There exists a non-trivial fixed point when $V_0<0$ as $X_+=X_-= \sqrt{- 
\frac{\alpha^2}{2V_0}}$.  Therefore, if $- \frac{\alpha^2}{2V_0}=1$, there 
is an asymptotically flat spacetime. In the following, we assume that 
$V_0=- \alpha^2/2 $ and we rewrite Eq.~(\ref{FRR4C}) as 
\begin{eqnarray} 
\label{FRR4D} X_+'&=& { -} \frac{\kappa^2 \alpha^2}{2} r X_+^2 X_-^{-1} 
\left(X_+^{-1} X_-^{-1} - 1 \right) \, , \nn
 X_-' &=& \frac{X_- \left( X_+ X_-^{-1} - 1\right)}{2r} \, . 
\end{eqnarray} 
Note also that $X_+'=0$ when $X_+ X_-=1$ and $X_-'=0$ when $X_+=X_-$. More 
precisely, the situation is summarized as follows.

{\begin{enumerate}

\item When both $X_+$ and $X_->0$, \begin{enumerate} \item if 
$X_->X_+^{-1}$, then $X_+'>0$; \item if $X_-<X_+^{-1}$, then $X_+'<0$; 
\end{enumerate} and \begin{enumerate} \item if $X_->X_+$, then $X_-'<0$; 
\item if $X_-<X_+$, then $X_-'>0$. \end{enumerate}

\item When $X_+<0$ and $X_->0$, then $X_+'>0$ and $X_-'<0$; \item When 
$X_+$, $X_-<0$, \begin{enumerate} \item if $X_->X_+^{-1}$, then $X_+'>0$; 
\item if $X_-<X_+^{-1}$, then $X_+'<0$. \end{enumerate} and 
\begin{enumerate} \item if $X_->X_+$, then $X_-'<0$; \item if $X_-<X_+$, 
then $X_-'>0$. \end{enumerate}

\item When $X_+>0$, we have $X_-<0$, $X_+'<0$, and $X_-'>0$. 
\end{enumerate} }

We also note that $X_+'\to 0$ as $X_+\to 0$. Furthermore, when $X_-\to 0$, 
then $X_+'\to { -} \infty$ $\left( X_+>0\right)$ and $X_+'\to { +} \infty$ 
$\left(X_+< 0\right)$. In the Schwarzschild-like solution with 
$g_{tt}g_{rr}=-1$, $X_+=1$ and $X_-$ vanishes at horizon. Therefore, the 
solution with $g_{tt}g_{rr}=-1$ does not exist in the model~(\ref{FRR4C}).

We now perturb the fixed points $X_\pm=1$ as described by $X_\pm = 1 + 
\delta X_\pm$. Equation~(\ref{FRR4C}) yields \begin{eqnarray} \label{FRR4E} 
\delta X_+' &=& \frac{\kappa^2 \alpha^2 r}{2} \left( \delta X_+ + \delta 
X_- \right) \, , \\ \delta X_-' &=& \frac{1}{2r} \left( \delta X_+ - \delta 
X_- \right) \, . \end{eqnarray} The eigenvalues $\lambda$ of the matrix 
\begin{equation} \hat{M} \equiv \left( \begin{array}{cc} \frac{\kappa^2 
\alpha^2 r}{2}
& \frac{\kappa^2 \alpha^2 r}{2} \\
\frac{1}{2r} & - \frac{1}{2r} \end{array} \right) \end{equation} are the 
roots of the secular equation \begin{equation} \label{FRR4F} \lambda^2 - 
\left( \frac{\kappa^2 \alpha^2 r}{2} - \frac{1}{2r} \right)  - 
\frac{\kappa^2 \alpha^2}{4} = \left( \lambda - \frac{\kappa^2 \alpha^2 
r}{2} \right) \left( \lambda + \frac{1}{2r} \right) =0 \, , \end{equation} 
hence both eigenvalues $ \kappa^2 \alpha^2 r/2 $ and $- 1/(2r) $ are 
negative and the equilibrium point $X_\pm=1$ is a saddle point. The 
negative eigenvalue $- 1/(2r) $ corresponds to the spacetime where $r$ 
becomes larger, $X_\pm\to 1$, therefore this geometry is asymptotically 
flat.

There is also a line of (non-isolated) fixed points at $X_-\to \infty$. To 
see this, we rewrite Eq.~(\ref{FRR4D}) as \begin{eqnarray} \label{FRR4G} 
X_+' &=& { -} \frac{\kappa^2 \alpha^2}{2} \, r X_+^2 X_-^{-1} 
\left(X_+^{-1} X_-^{-1} - 1 \right) \, , \nn \left( X_-^{-1} \right)' &=& - 
\frac{X_-^{-1} \left( X_+ X_-^{-1} - 1\right)}{2r} \, . \end{eqnarray} 
Then, at $X_-^{-1}=0$ we find $X_+'= \left( X_-^{-1} \right)' = 0$. In 
order to investigate the stability, we consider the perturbation $\delta 
X_-^{-1}$ of $X_-^{-1} $ by fixing $X_+$. The second equation in 
(\ref{FRR4G}) reduces to \begin{equation} \label{FRR4H} \left( \delta 
X_-^{-1} \right)' \sim \frac{\delta X_-^{-1} }{2r} \, . \end{equation} The 
eigenvalue $ 1/(2r)$ is always positive.

Then there are several types of solutions. By fine-tuning the initial 
condition, one could find a trajectory which begins near $X_-\to +\infty$ 
and approaches the fixed point $X_\pm = 1$. There could also be 
trajectories beginning near $X_-\to +\infty$, crossing the line $X_+=X_-$, 
and approaching to $X_+\to +\infty$. The details are not particularly 
illuminating and will not be reported, however, the main point is that the 
existence of time-dependent and spherically symmetric solutions in pure 
$f(R)$ gravity is determined.

\section{Conclusions} \label{sec:7} \setcounter{equation}{0}

Many of our examples are plagued by the presence of a ghost, a phantom 
scalar field with negative kinetic energy. Such phantom fields suffer from 
instability and should not exist: however, from time to time cosmological 
observations argue in favour of a phantom equation of state of the cosmic 
quintessence fluid \cite{Caldwell:1999ew,Caldwell:2003vq,Nojiri:2005sr, 
Nojiri:2005sx,Melchiorri:2002ux}. Should such claims persist, one would be 
forced to take the phantom phenomenology more seriously, perhaps not as 
signalling a true phantom field, but as a phantom-like phenomenon arising 
in a theory that is fundamentally ghost-free (this happens occasionally in 
modified gravity, for example in $f(R)$ gravity, whose gravitational sector 
is ghost-free \cite{Sotiriou:2008rp,Nojiri:2017ncd,Nojiri:2006ri}). In this 
case, a model with a phantom field would be a mimicker of a more viable 
theory and the black hole solutions described in the previous sections 
would then become more interesting. These solutions include a static 
apparent horizon in a time-dependent black hole geometry, in which the 
black hole accretes scalar fluid and changes its temperature, while keeping 
its horizon unchanged. The first of our three ansatzes produces genuine 
time-dependent black holes with apparent horizons scaling in time with the 
Misner-Sharp-Hernandez mass (this horizon will look the same to all 
observers associated with a spherically symmetric foliation 
\cite{Faraoni:2016xgy}, but not to those moving with respect to the former 
in such a way that this symmetry is broken 
\cite{Wald:1991zz,Schnetter:2005ea}, for example by a Lorentz boost along a 
non-radial direction).

Thus far, solving directly the field equations of various theories has not 
produced many physically reasonable solutions describing dynamical black 
holes; naked singularities and wormholes are much more common 
\cite{Vbook,Fisher:1948yn,Bergmann:1957zza,Janis:1968zz, 
Buchdahl:1972sj,Wyman:1981bd,Bronnikov:1973fh,Campanelli:1993sm, 
Vanzo:2012zu,Faraoni:2018mes,Fonarev:1994xq,Kastor:2016cqs,
Faraoni:2017afs,Banijamali:2019gry,Faraoni:2017ecj}. Another problem is 
Faraoni:that some
of these analytical solutions are cumbersome when expressed in terms of the 
areal radius ({\em e.g.}, 
\cite{Clifton:2004st,Clifton:2006ug,Faraoni:2007es,Gao:2008jv})  and often 
the apparent horizons cannot be located analytically, or the analytical 
expressions providing them are not explicit \cite{Vbook}. Overall, 
designing time-dependent black holes is difficult and we reverse-engineered 
the coupling functions of theories of gravity that admit prescribed 
apparent horizons as their solutions. Simpler analytical dynamical black 
holes will be searched for in future work.

\section*{Acknowledgments}

We thank a referee for useful comments leading to an improved 
presentation. This work was supported by MINECO (Spain), project 
PID2019-104397GB-I00 and PHAROS COST Action (CA16214) (S.D.O.), by JSPS 
Grant-in-Aid for Scientific Research (C) No. 18K03615 (S.N.), and by the 
Natural Sciences \& Engineering Research Council of Canada, Grant 
No.~2016-03803 (V.F.). The work by S.D.O. was partially supported by the 
Ministry of Science and High Education of Russia, project No. 
FEWF-2020-003.

\appendix \section{} \label{AppendixA} \setcounter{equation}{0} 
\renewcommand{\theequation}{A.\arabic{equation}}

When the metric can be regarded as static, or its time-dependence can be 
neglected, we consider the line element \begin{eqnarray}
 ds^2 &=& - P(r) \left( r - r_0 \right) dt^2 + \frac{dr^2}{ P(r) \left( r - 
r_0 \right) } \nn
&\, & + r^2 \left( d\vartheta^2 + \sin^2\vartheta \, d\varphi^2 \right)\,
. \end{eqnarray} We assume that $P(r)$ is positive everywhere and is a 
sufficiently smooth function of $r$ near the horizon $r=r_0$, and that it 
can be approximated by a constant, $P(r)\sim P(r_0)$. We then introduce a 
new coordinate $\rho$ by \be d\rho = \frac{dr}{\sqrt{ P(r_0) \left( r - r_0 
\right) }}\, , \ee that is, \be \rho = 2 \sqrt{\frac{r - r_0}{P(r_0)}}\, . 
\ee We also Wick-rotate the time coordinate as $t\to i\tau$. Then, we 
obtain the Euclidean metric \be ds^2 = \frac{P\left(r_0\right)^2}{4} \rho^2 
d\tau^2 + d\rho^2
+ r(\rho)^2 \left( d\vartheta^2 + \sin^2\vartheta \, d\varphi^2 \right)\,.  
  \ee In order to avoid the conical singularity at $\rho=0$, we need to 
  impose the periodicity on $\tau$, \be \frac{P\left(r_0\right)}{2} \tau 
  \sim \frac{P\left(r_0\right)}{2} \tau
+ 2 \pi \, . \ee For the finite temperature formalism in the path-integral, 
  the periodicity $\frac{4\pi}{P\left(r_0\right)}$ corresponds to the 
  inverse of the temperature \be T = \frac{P\left(r_0\right)}{4\pi}\, . \ee 
  In the case of the metric given by~(\ref{Ex2a}), we find \be \ P(r_0) = 
  \frac{t_0}{t r_0}\, , \ee and we obtain the expression~(\ref{T1e2}). $t$ 
  and $T$ are the Kodama time and Kodama temperature, respectively.


\section{} 
\label{AppendixB} 
\setcounter{equation}{0} 
\renewcommand{\theequation}{B.\arabic{equation}}

The action for Einstein gravity with a scalar field doublet is 
\begin{align} 
\label{II8} 
S_{(\phi\chi)} = \int d^4 x \sqrt{-g} & \left\{ 
\frac{R}{2\kappa^2}
 - \frac{1}{2} \, A (\phi,\chi) \, \partial_\mu \phi \partial^\mu \phi 
   \right.\nn
&\left.
 - B (\phi,\chi) \, \partial_\mu \phi \partial^\mu \chi \right. \nn & 
   \left.  - \frac{1}{2} \, C (\phi,\chi) \, \partial_\mu \chi \partial^\mu
\chi - V (\phi,\chi)\right\}\, . 
\end{align}

In the static case where $B=0$, the action~(\ref{II8}) reduces to 
\begin{eqnarray} \label{SSII2} S_{(\phi\chi)} &=& \int d^4 x \sqrt{-g} 
\left\{ \frac{R}{2\kappa^2}
 - \frac{1}{2} \, A(\chi) \, \partial_\mu \phi \partial^\mu \phi 
   \right.\nonumber\\
&\, & \left.
 - \frac{1}{2}\, C(\chi) \, \partial_\mu \chi \partial^\mu \chi - V (\chi)  
   \right\}\, . \end{eqnarray} We now redefine the scalar field $\chi$ as 
   \begin{equation} \label{SII3} \tilde\chi = \int d\chi \sqrt{ \left| C 
   (\chi) \right|} \end{equation} and we rewrite the action~(\ref{SSII2})  
   as \begin{align} \label{SII4} S_{(\phi\chi)} =& \int d^4 x \sqrt{-g} 
   \left\{ \frac{R}{2\kappa^2}
 - \frac{1}{2} \, \tilde A \left( \tilde\chi\right) \, \partial_\mu \phi 
   \partial^\mu \phi \right.\nn
& \left.
 - \frac{1}{2} \, \mathrm{sgn} \left( C \right)\, \partial_\mu \tilde\chi 
   \partial^\mu \tilde\chi
 - \tilde V \left( \tilde\chi \right)\right\}\, , \nn &\nn & \tilde A 
   \left( \tilde\chi \right) \equiv A \left( \chi\left( \tilde\chi
\right)\right) \,, \quad \quad \tilde V \left( \tilde\chi \right) \equiv V 
\left(\chi\left( \tilde\chi \right)\right) \, , \end{align} where 
\begin{equation} \label{SII5} \mathrm{sgn} \left( C \right) \equiv \left\{ 
\begin{array}{ll}
+1 & \ \mbox{when}\ C> 0 \, , \\
 -1 & \ \mbox{when}\ C< 0 \, . \end{array} \right. \end{equation} We now 
  consider the relation between the model (\ref{SSII2}) and $F(R)$ gravity, 
  or models similar to $F(R)$ gravity.  For this purpose, assume that 
  $C>0$, that is, $\mathrm{sgn} \left( C \right)=1$. Under the scale 
  transformation $g_{\mu\nu}\to \e^\rho g_{\mu\nu}$, the Ricci scalar 
  transforms as \begin{equation} \label{E4} R \to \left(R - 3\Box \rho - 
  \frac{3}{2} \, \partial^\mu \rho \partial_\mu \rho \right)\e^{-\rho} 
  \end{equation} and the action (\ref{SSII2}) with $\mathrm{sgn} \left( C 
  \right)=1$ transforms as \begin{align} \label{SII6} S_{(\phi\chi )} = 
  \int d^4 x \sqrt{-g} \, \e^{2\rho} & \left\{ \frac{1}{2\kappa^2} \left( R
  - 3\Box \rho - \frac{3}{2} \, \partial^\mu \rho \partial_\mu \rho 
    \right)\e^{-\rho} \right.\nn
&\left.
 - \frac{1}{2} \, \tilde A \left( \tilde \chi \right) \e^{-\rho} \, 
   \partial_\mu \phi \partial^\mu \phi \right. \nn
& \left.
 - \frac{1}{2}\, \e^{-\rho}\, \partial_\mu \tilde\chi \partial^\mu 
   \tilde\chi - \tilde V \left( \tilde\chi \right)\right\}\, . \end{align} 
   Then, the choice $\rho = \tilde\chi / \sqrt{3}$ yields \begin{eqnarray} 
   \label{SII7} S_{(\phi\chi )}& &= \int d^4 x \sqrt{-g} \left\{ 
   \frac{1}{2\kappa^2} \, \e^{\frac{\tilde\chi}{\sqrt{3}}} \, R
 - \frac{1}{2} \, \tilde A \left( \tilde \chi \right) \, 
   \e^{\frac{\tilde\chi}{\sqrt{3}}} \, \partial_\mu \phi \partial^\mu \phi 
   \right.\nonumber\\
& \, & \left. - \e^{\frac{2\tilde\chi}{\sqrt{3}}} \, \tilde V \left( 
\tilde\chi \right)\right\}\, . 
\end{eqnarray} 
By varying with respect to $\tilde\chi$, one obtains the algebraic 
equation 
\begin{align} 
\label{SII8} 
0 =& \frac{1}{\sqrt{3}} \left\{ \frac{1}{2\kappa^2} \, 
\e^{\frac{\tilde\chi}{\sqrt{3}}} \, R
- \frac{1}{2} \, \tilde A \left( \tilde \chi \right) \, 
 \e^{\frac{\tilde\chi}{\sqrt{3}}} \, \partial_\mu \phi \partial^\mu \phi
 - 2 \, \e^{\frac{2\tilde\chi}{\sqrt{3}}} \, \tilde V \left( \tilde\chi 
   \right)\right\} \nn
& - \frac{1}{2} \, \tilde A' \left( \tilde \chi \right) \, 
   \e^{\frac{\tilde\chi}{\sqrt{3}}} \, \partial_\mu \phi \partial^\mu \phi
 - \e^{\frac{2\tilde\chi}{\sqrt{3}}} \, \tilde V' \left( \tilde\chi \right)  
   \, , 
\end{align} 
which can, in principle, be solved with respect to $\tilde\chi$ as 
$\tilde\chi= \tilde\chi \left( R, \partial_\mu \phi  \partial^\mu \phi 
\right)$. Then by substituting the expression of $\tilde\chi \left( R, 
\partial_\mu \phi \partial^\mu \phi \right)$ into  the action 
(\ref{SII7}), we obtain an action which is, in a sense, similar to the 
action of $F(R)$ gravity, 
\begin{equation} \label{SII9} 
S_{( \phi\chi )} = \int d^4 x \sqrt{-g} \, F \left( R, \partial_\mu \phi  
\partial^\mu \phi \right) \, , 
\end{equation} 
where 
\begin{eqnarray} 
F \left( R, \partial_\mu \phi \partial^\mu \phi \right) \equiv \, 
\frac{1}{2\kappa^2} \, \e^{\frac{\tilde\chi\left( R, \partial_\mu \phi  
\partial^\mu \phi \right)}{ \sqrt{3}}} \, R \nonumber\\ 
\quad -  \frac{1}{2}\, \tilde A \left( \tilde \chi\left( R, \partial_\mu 
\phi \partial^\mu \phi \right) \right) \, \e^{\frac{\tilde\chi\left( R,  
\partial_\mu \phi \partial^\mu \phi \right)}{ \sqrt{3}}} \, \partial_\mu 
\phi \partial^\mu \phi \nonumber\\ 
\quad - \e^{\frac{2\tilde\chi\left(  R, \partial_\mu \phi \partial^\mu 
\phi \right)}{\sqrt{3}}} \, \tilde V  \left( \tilde\chi\left( R, 
\partial_\mu \phi \partial^\mu \phi \right)   \right) \, . 
\end{eqnarray} 
Therefore, even in the static case, Einstein gravity with two scalar 
fields cannot be rewritten in a simple $F(R)$ gravity form, but we obtain 
a rather complicated model instead.

In the time-dependent case it might be possible, in principle, to rewrite 
the model in a way similar to the form of $F(R)$ gravity, although the 
situation is not simple since the model obtained in this way assumes very 
complicated forms.

The physical nature of analytic solutions of the field equations may change 
radically when one goes from static to time-dependent solutions. In GR with 
a single scalar field, static black holes are Schwarzschild, or else the 
Fisher solution, which describes a naked singularity, is obtained for a 
single, static, asymptotically flat scalar field. However, when the 
geometry and the free scalar field are allowed to be time-dependent, one 
obtains the Husain-Martinez-Nu\~nez solution, which describes a 
cosmological black hole for part of its history \cite{Husain:1994uj}. 
Similar conclusions are reached by introducing an exponential potential for 
the scalr field, which produces the Fonarev solution \cite{Fonarev:1994xq}.

When one contemplates time-dependent, asymptotically FLRW solutions of the 
Einstein equations with a fluid, one finds the McVittie metric 
\cite{McVittie:1933zz}. The latter cannot be generated as a scalar field 
solution of the Einstein equations, or as a solution of scalar-tensor 
gravity. However, it is an exact solution of cuscuton theory, a special 
case of Ho\u{r}ava-Lifschitz gravity \cite{Abdalla:2013ara}, which is the 
only form of $k$-essence that admits McVittie solutions 
\cite{Abdalla:2013ara}.  The McVittie spacetime is also an solution of 
shape dynamics \cite{Gomes:2010fh}, and of $f(T)$ gravity (where $T$ is the 
torsion scalar \cite{Bejarano:2017akj}). $f(R)$ gravity with a single 
scalar is essentially equivalent to GR with two scalars and, while one does 
not expect much difference from the case of GR with a single scalar in the 
static case, one expects the time-dependent, asymptotically FLRW situation 
to produce more analytic solutions which contain dynamical black hole 
apparent horizons for part of the spacetime history, as in the case of the 
Husain-Martinez-Nu\~nez \cite{Husain:1994uj} and Fonarev 
\cite{Fonarev:1994xq} solutions.

\end{document}